\documentclass{article}
\usepackage{emulateapj,pstricks,apjfonts}
\usepackage{lscape}

\def\lae{\mathrel{<\kern-1.0em\lower0.9ex\hbox{$\sim$}}}
\def\gae{\mathrel{>\kern-1.0em\lower0.9ex\hbox{$\sim$}}}

\def\bfs{\mbox{blank field sources}}

\def\1wga{\mbox{1WGA~J1226.9$+$3332}}
\def\fxv{\mbox{$f_X/f_V$}}
\def\cgs{\mbox{erg cm$^{-2}$ s$^{-1}$}}
\begin{document}

\submitted{Accepted version. To appear in ApJ main journal}

\lefthead{{\it ROSAT} blank field sources {\sc i}}
\righthead{Cagnoni et al.}

\title{{\it ROSAT} Blank Field  Sources {\sc i}: Sample Selection and Archival Data}

\author{I. Cagnoni\altaffilmark{1,2,3},  M. Elvis\altaffilmark{3},
D.W. Kim\altaffilmark{3},  F. Nicastro\altaffilmark{3} \and  A. 
Celotti\altaffilmark{2} }
\affil{$^{1}$ Dipartimento di Scienze, Universit\`a dell'Insubria, Como, Italy}\affil{$^{2}$ SISSA, Via Beirut 4 -34138, Trieste, Italy}
\affil{$^{3}$ Harvard-Smithsonian Center for Astrophysics, 60 Garden Street, 
Cambridge, MA 02138, USA}
\footnote{email: Ilaria Cagnoni: ilaria.cagnoni@uninsubria.it}

\begin{abstract}
We have identified a population of `\bfs ' (or `blanks') among 
the {\em ROSAT} bright unidentified 
X-ray sources with faint optical counterparts.
The extreme X-ray over optical flux ratio of blanks is not compatible with 
the main classes of X-ray emitters  except for extreme BL Lacertae objects.\\ 
From the analysis of {\em ROSAT} archival data we found no indication 
of variability and evidence for only three sources, out of 16,  
needing absorption in excess of the Galactic value.
We also found evidence for an extended nature for only
one of the 5 blanks with a serendipitous HRI detection; this source
(1WGA~J1226.9$+$3332) was confirmed as a z=0.89 cluster of galaxies.
Palomar images reveal  the presence of a red
($O-E \geq$ 2) counterpart in the X-ray error circle for 6  blanks.
The identification process brought to the discovery of another high z 
cluster of galaxies, one (possibly extreme) BL Lac, two 
ultraluminous X-ray sources in nearby galaxies and 
two apparently normal type~1 AGNs.
These AGNs, together with 4 more AGN-like 
 objects  seem to form a well defined group:
they present unabsorbed X-ray spectra but red Palomar counterparts.
We discuss the possible explanations for the discrepancy between
the  X-ray and optical data, among which: a
suppressed big blue bump emission,
an extreme  dust to gas ($\sim 40-60$ the Galactic ratio),
 a high redshift ($z \geq 3.5$) QSO nature,  an atypical
 dust grain size distribution and a dusty warm absorber.
These AGN-like blanks seem to be the bright (and easier to study)  
analogs of the sources which are
 found in deep {\em Chandra} observations. 
Three more blanks have a still unknown nature. 
\end{abstract}


\section{Introduction}

The X-ray sky is not as well known as sometimes thought. 
There might exist classes of quite common sources, comprising from 
one to a few percent 
of the  high Galactic latitude population, which are currently 
thought to be ``rare'' 
because of the difficulty of finding them.
We are actively searching for such `minority' populations (Kim \& Elvis 1999).
We have found an interesting high fraction of extreme X-ray loud sources
($\sim 7-8$\%) among the {\em ROSAT}
 high Galactic latitude sources at all fluxes: 
(this paper, Bade et al. 1995; Schmidt et al. 1998).
Here we designate as  `blank field sources' or `blanks' all the bright 
{\em ROSAT} PSPC (Position Sensitive Proportional Counter) X-ray sources 
($f_X > 10^{-13}$ erg cm$^{-2}$ s$^{-1}$) with no optical counterpart 
on the Palomar Observatory Sky Survey (POSS) (to O=21.5\footnote{O-band effective wavelength = 4100 \AA ; passband = 1100 \AA}) within
their $39^{\prime \prime}$ (99\%) radius error circle.
For comparison, at these X-ray fluxes a normal type 1 AGN should appear on the
POSS some 3.5 magnitudes brighter ($O \sim 18$).
The nature of these sources has never been systematically investigated before,
and their nature is still mysterious.
We decided to select these sources and to study them to understand their
nature  because of the important cosmological and astrophysical 
consequences that could derive from their identification.\\
The outline of this paper is the following: 
we introduce  {\em ROSAT} blank field sources,
review the open possibilities for their nature and the important 
consequences that could 
derive from the identification of these sources  in \S ~2; 
we present the selection criteria used for the definition 
of the sample in \S ~3, while \S ~4, 6 and 7 focus
 on the analysis of {\em ROSAT}, {\em ASCA} and radio 
archival data.
\S ~7 contains detailed information on  the sources
and  the discussion, while in \S ~8 we compare the results of this work 
with the findings of other surveys. A summary is presented in \S ~9.
We will use H$_0$=75~km~s$^{-1}$~Mpc$^{-1}$ and $q_0 = 0.5$;
errors   represent 1$\sigma$ confidence levels, unless 
explicitly stated otherwise.

\section{Potential Classes of Blank Field Sources}
\label{blanks_nature}

Blank field sources are extreme X-ray loud objects with $f_X/f_V > 10$\footnote{We follow the $f_X/f_V$ definition of Maccacaro et al. (1988)}.
The Maccacaro et al. (1988)  nomograph allows us to compare this  $f_X/f_V$ 
with that typical of other classes of objects.
Normal quasars and AGN have 0.1$< f_X/f_V <$10;  normal galaxies have only 
$10^{-2}< f_X/f_V < 10^{-1}$; while a high luminosity cluster of galaxies
($L_X=10^{45}$ erg~s$^{-1}$) would have a first ranked elliptical
with $L_{gal}=10L^{\ast} (=10^{11.5}L_{\odot}$) giving $f_X/f_V \sim 8$.
The only known class of X-ray emitting objects that can reach such extreme \fxv \/ 
are BL Lacertae objects, which can have $f_X/f_V$ up to $\sim 35$ 
(Maccacaro et al. 1988).

There are however several types of hypothesized objects that would appear as
\bfs . We briefly describe these below.
 
\subsection{Extreme BL Lacertae objects}

``Normal'' BL Lacs are expected for $f_X/f_V < 35$ (e.g. Maccacaro et al. 1988), but given the range of $f_X/f_V$ spanned by blanks
more extreme BL Laces, with the SED peaking at energies
higher than the {\em ROSAT} band are needed.
This peak could be either the low energy (synchrotron)
or the high energy  (inverse Compton) one shifted appropriately.
Fossati et al (1998) show that these two peaks do shift systematically 
with luminosity.
In the first case, these indeed  would be faint BL Lacs, with the 
synchrotron peak at 
$1 < E \leq 10$ keV; some such extreme BL Lacs have been found recently by
e.g. Costamante et al. (2001).
No examples of blazars with an inverse Compton peak lying in the 1-10 keV 
range are known so far, but the study 
of {\em ROSAT} \bfs \/ could potentially discover them.

\subsection{Isolated Neutron Stars (INS) or Isolated Black Holes (IBH)}
\label{sec_ins} 

Blank field sources could be identified with old isolated neutron stars (INS) 
or isolated black holes (IBH), which could be observable due to accretion from the interstellar medium (``accretors''). 
INS can also be visible in the soft X-ray band when they are young and hot 
(``coolers'') due to thermal emission of their hot surface 
(e.g. Treves et al. 2000; Popov et al. 2002).
Old INS are expected to have $f_X/f_V > 100-1000$ (Maoz, Ofek \& Shemi, 1997).
and extremely soft X-ray spectra and emission in the UV. 
The active search for INS has produced 7 strong candidates 
(Stocke et al. 1995; Walter, Wolk \& Neuhauser 1996; 
Haberl et al. 1997; Haberl, Motch \& Pietsch 1998;
Schwope et al. 1999; Motch et al. 1999, Haberl, Pietsch \& Motch 1999;
Zampieri et al. 2001), but their very nature is not yet clear.
The large initial velocity found for one of these sources 
($\sim 200$ km s$^{-1}$ found by Walter (2001) for RX~J1856-3754)
supports the ``cooler'' origin, while no proof of the existence of 
``accreting'' old INS has been found. 
Being nearby INS and IBH ($\sim 100$ pc) 
will be quasi-isotropically distributed on the sky. 
The predicted number counts for old INS are highly uncertain, ranging from 
$\sim 1/$deg$^2$ (Madau \& Blaes 1994) to much smaller values 
(e.g. Colpi et al. 1998).
At the predicted number of old INSs strongly depends on the properties of 
their progenitors (young neutron stars in the pulsar phase), 
such as the velocity distribution and the magnetic field decay,
 the spatial density, or an upper limit on it, 
can be used to constrain neutron stars physics (e.g. Colpi et al. 1998;
Livio, Xu \& Frank, 1998; Popov et al. 2000).

\subsection{Peculiar AGNs}

Normal type 1 AGNs with $f_X > 10^{-13}$ \cgs \ 
have $O \sim 18$. Obscured AGNs could appear as blanks 
if their O-band absorption is $A_O > 3.5$~mag, which corresponds to 
a local $N_H > 4 \times 10^{21}$~cm$^{-2}$.
Possibilities for absorbed AGNs 
with $f_X >10^{-13}$ \cgs in the  {\em ROSAT} PSPC are:

\subsubsection{Quasar-2s}

QSO-2s would be  high luminosity, high redshift, heavily
obscured quasars, the bright analogs of the well known Seyfert~2s at
$z >> 0.5$. 
QSO-2s and Low Mass Seyfert-2 (LMSy2) galaxies are the ``missing links''
of the unifying models of AGN (Urry \& Padovani 1995) and their existence 
is still debated (e.g. Halpern, Turner \& George 1999).
A large number of QSO-2s is also postulated by the 
models of synthesis of the X-ray background above 2 keV to reproduce the 
observed number counts at bright fluxes (Comastri et al. 2001, 
Gilli, Salvati \& Hasinger 2001). 
Thus  a knowledge of 
their number density is fundamental to our understanding of the X-ray sky
and should be the site where a large fraction of the energy of the 
 universe is generated  (Fabian \& Iwasawa 1999; Elvis, 
Risaliti \& Zamorani, 2002).

In spectral terms we would expect: (a) strong optical reddening;
(b) soft X-ray absorption; 
(c) recovery of the X-ray intrinsic emission in the
PSPC  energy range. 
QSO-2s should then have flat PSPC spectra and an observed 
$f_X/f_V$ enhanced by the redshift K-correction.
For moderate absorbing columns ($N_H \sim 10^{22} - 10^{23}$ cm$^{-2}$) 
QSO-2s would be visible in the PSPC because their redshift has the 
effect of lowering the observed absorption; for higher column densities
the direct emission could be completely absorbed, but a scattered component 
could still allow the source detection.

In order to quantify the expected X-ray properties, we simulated these
 two  QSO-2 scenarios assuming, 
for the moderately absorbed case,  a rest frame absorbed  ($N_H = 5 \times
10^{22}$ cm$^{-2}$)  power law model ($\Gamma = 1.7$)
at z=1.0 with an unabsorbed 
 $L_{(0.5-2.4 {\rm keV})} = 1.8 \times 10^{45}$ erg s$^{-1}$.
We also included a neutral iron line at $E_{Fe}=6.4$ keV and
equivalent width of 0.15 keV.
The simulated QSO-2 is detected by the {\em ROSAT} PSPC with
a count rate (CR) of $\sim 0.01$ counts s$^{-1}$ corresponding to a WGACAT 
(0.05-2.5 keV) flux
of $2.8 \times 10^{-13}$ \cgs .\\
For the highly absorbed case we model the spectrum as the sum of an absorbed spectrum and a scattered component.\\
$f(E)= e^{-\sigma N_H} E^{-\Gamma} + K E^{-\Gamma}$ at z=1.0\\
where  $\sigma$ is the neutral H cross section for absorption, 
$N_H \geq 10^{24}$ cm$^{-2}$ in the rest frame,
$\Gamma = 1.7$ 
and $K=0.1$ indicates a 10\% scattered component.
With the assumed values of $N_H$ the direct component is completely blocked 
by the absorbing material. 
A QSO with an unabsorbed luminosity of $L_{(0.5-2.4 {\rm keV})} \geq 10^{46}$ 
erg s$^{-1}$ would be detected as scattered light in the {\em ROSAT} PSPC band with a WGACAT
flux of $\geq  10^{-13}$ \cgs .\\
Both would thus be included among \bfs .

\subsubsection{Low Mass Seyfert~2}

LMSy~2 are 
Seyfert~2 galaxies powered by a relatively low mass obscured black
hole ($M_{bh} \sim 10^7 M_{\odot}$). 
These would  be the obscured analogs of the Narrow Line Seyfert 1s.
Their low mass would shift the peak of the big blue bump (BBB) 
disk emission ($T_{peak} \sim M_{bh}^{1/4}$) around 100 eV.
As a result their  spectra would be enhanced in the soft X-ray
band and suppressed in the optical.
RE~1034+396 (Puchnarewicz et al. 1995; 2001) 
shows that this does indeed occur in NLSy~1.
We simulated the LMSy2 case assuming an absorption of $5 \times
10^{22}$ cm$^{-2}$, a thermal component of  kT = 150 eV, a power
law with $\Gamma = 2.2$.
We normalized the model to have a ratio between the
thermal and power law component matching what observed in Narrow Line
Sy~1s (Nicastro, Fiore \& Matt, 1999) and to have an unabsorbed luminosity
at z=0.002 of $L_{(0.5- 2.4 {\rm keV})} = 4 \times 10^{42}$ erg~s$^{-1}$.
The simulated LMSy2 is detected in the {\em ROSAT} PSPC with a count rate
of 0.015 counts s$^{-1}$ corresponding to a 0.5 -2.4 keV flux of 
$2.4 \times 10^{-13}$ \cgs . 
Thus obscuration, even by large amounts of molecular matter and cold gas 
($N_H = 10^{22}-10^{23}$ cm$^{-2}$), would still allow the detection of 
the  soft X-ray source emission because it is so intrinsically strong, 
but would harden the observed X-ray spectra.

\subsubsection{AGNs with no big blue bump}

Virtually all type~1 AGNs have optical-ultraviolet 
spectra dominated by the BBB component from the accretion disk 
(Malkan \& Sargent, 1982; Elvis \& Lawrence, 182; Elvis et al. 1994)
There are however   radiatively inefficient
 mechanisms for  accretion. An example of this are 
`advection dominated accretion flows' (ADAFs)
(e.g. Narayan \& Yi, 1994), where 
 most of the energy is stored in the gas and advected toward 
the hole and only a small fraction is radiated.
The result is a suppression of the optical-UV-EUV BBB emission 
(see Figure~1 of Narayan et al. (1998) for a comparison of 
ADAF emission 
to the standard Shakura-Sunyaev (1973) thin disk emission).

Radiatively inefficient accretion has been proposed to account for 
 the spectra of Galactic black holes 
X-ray binaries (BHXBs)
and Sagittarius~A* (Narayan et al. 1998).
Finding radiatively inefficient accretion among blanks
could shade light on the physics and history of the accretion process in AGN and possibly  strengthen the AGN-BHXB connection.

\subsection{Unusual clusters of galaxies}
\label{nature_clusters}

\subsubsection{High Redshift Clusters of Galaxies}

At high redshifts, $z \geq 0.4$, the 4000 \AA \/ break in 
the spectrum of normal  galaxies is shifted to wavelengths longer than 
the POSS O band.
Any high z cluster of galaxies will then be fainter than O=21.5, and have 
$f_X/f_V > 10$.
Using `blanks' to locate high z clusters 
efficiently is a promising path to pursue.
With a WGACAT (0.05-2.5 keV) $f_X >10^{-13}$ \cgs \/ and assuming Galactic absorption,
all such clusters will have instrinsic $L_X \geq 10^{44}$ erg s$^{-1}$.
The space density of such high luminosity, and by implication massive,
 high z clusters of galaxies is a  crucially important tool in cosmology.
Their distribution and evolution is  fully determined by the spectrum
of primordial perturbations and cosmological parameters $\Omega_0$ and 
$\Lambda$ (e.g. Press \& Schechter 1974). 
In particular,  models of low $\Omega$ universe (with or without cosmological 
constant)  predict a higher density 
of massive clusters  at high redshifts than  the high $\Omega$ models
(e.g. Henry 2000, Borgani \& Guzzo 2001).\\
Only few bright high redshift clusters have been found so far 
(e.g. in the EMSS, Gioia \& Luppino 1994; in the RDCS Rosati et al. 1999,
 Della Ceca et al. 2000; in the WARPS Ebeling et al. 2000) and only 11
 of them at $z>0.5$ have a measure of their temperature
(e.g. Della Ceca et al. 2000, Cagnoni et al. 2001a, Stanford et al. 2001).\\
Since the statistics is scanty, even a few
 high redshift clusters can help 
determining the evolution of the clusters X-ray luminosity function 
(e.g. Rosati et al., 1998) and any search for 
evolution in the luminosity-temperature ($L_X-T$) relation, 
 related to the physical mechanisms of cooling and heating in the central
 cluster region (e.g. Tozzi \& Norman, 2001).
Since the statistics is very scanty, adding one more high z high L cluster 
would greatly improve such studies.

\subsubsection{Dark clusters}

Dark clusters are unusual clusters of galaxies with extreme high mass to light ratio.
They have been searched for as ``missing lens'' in gravitational
lenses searches (e.g. Hattori et al 1997) and one of such objects
was found in  the lensed system MG2016$+$112.
The lensing mass did not show up in deep optical/IR imaging, while 
follow-up X-ray observations detected a $z=1$ cluster (Hattori et al. 1997).
Although {\em Chandra}  and deeper optical searches
found that the properties of the cluster AXJ2019$+$1127 are not so extreme as previously thought (Chartas et al. 2001), if dark clusters exist,
they could be found among blanks.

\subsubsection{Failed Clusters}

A failed cluster would be a large overdensity of matter in the form 
of diffuse gas only. Galaxies have to form from clouds of 
gas with cooling time shorter than the Hubble time.
This implies a protogalactic mass $<10^{12} M_{\odot}$ 
(e.g. White \& Frenk 1991 and references therein); in principle more massive 
clouds, with a longer cooling  might  have never 
collapsed to form stars in appreciable number. The result would be a large 
cloud of gas with no visible galaxies.\\
Their possible existence and number-density  
 is connected to the 
efficiency of galaxy formation, which potentially constrains 
a number of issues such as: 
the spectrum of the fluctuations that gave rise to galaxies, 
the baryonic or non-baryonic nature of dark-matter, 
the value of $\Omega$ and 
the relative distribution of dark and baryonic matter 
(Tucker, Tananbaum \& Remillard 1995).

The discovery of even only one failed cluster among our blanks would pose 
serious difficulties for the hierarchical clustering formation, 
which does not predict their existence.

\subsection{Ultraluminous X-ray sources and binary systems in nearby galaxies}

Ultraluminous X-ray sources (ULX) are objects observed in nearby galaxies,
usually off the galactic center,
with luminosity exceeding the Eddington limit for isotropic emission of a 
1.4 $M_{\odot}$ star (i.e. $L > 2 \times 10^{38}$ up to $\sim 10^{40}$,
e.g. Fabbiano 1898; Makishima et al. 2000; Zezas et al. 2002; Colbert \& Ptak 2002 and references therein).
The ULX nature is still not  understood and the most plausible explanations are:
sub-Eddington binaries involving  massive black holes ($\sim 100 M_{\odot}$) or binary systems with emission collimated toward our line of site (King et al. 2001) or relativistically beamed systems (microquasars).
These models imply an high $\fxv$ because the optical emission is 
related to the companion star.

Also X-ray binaries and cataclismic variables are known to display high 
X-ray over optical flux ratios and could be candidate blanks if
present in an uncrowded optical region of a nearby galaxy.

\subsection{Extreme variables}

\subsubsection{X-ray Bursts}

X-ray bursts are short lived  X-ray flares, which may be
 afterglows of  Gamma Ray Bursts (GRBs).
Popular GRBs scenarios such as 
binary coalescence of 
compact stars (e.g. Janka \& Ruffert, 1996) or collapsars 
(e.g. Hartmann \& MacFadyen, 1999) predict collimated flows, 
which should lead to collimated bursts and afterglow emission.
If afterglows turn out to be less beamed than GRBs, then we expect to 
find a higher rate of afterglows than GRBs depending on the 
beaming angles.
This possibility can be tested with a search for afterglows fortuitously 
detected during {\em ROSAT} PSPC pointings (e.g. Greiner et al. 2000 for 
a systematic search in the {\em ROSAT} All-Sky Survey, 
RASS).
The {\em ROSAT} lightcurves of afterglows could also
constrain   the X-ray flux decaying rate, which can be compared  with
the power law slope of $\sim 1.0 -1.5$ seen for the brighter {\em Beppo-SAX}
afterglows (Frontera et al. 2000).

\subsubsection{X-ray flares from normal galaxies}

Another class of variable sources are normal galaxies which exhibit
X-ray flares, probably related to the tidal distruption of a star near 
the central black hole (e.g. Bade, K., D., 1996).  
In the X-ray high-state,
these galaxies exceed their quiescent \fxv \/ by up to several orders of
magnitude. Timescales for the duration of the
X-ray high-state are not yet well constrained; they may range
from months to years.

\subsubsection{Strongly variable sources}

Blanks candidates also include variable sources, 
such as blazars or cataclismic variables, which were 
caught in outburst during  the 
{\em ROSAT} observations, but were quiescent in the POSS epoch, 
some 40 years earlier.

\subsection{Summary}

Quite possibly the class of \bfs \/ contains examples of several, or
all, of these objects. The possibility of identifying blanks with sources
listed above makes the study of this minority population
a subject of primary importance in cosmology and astrophysics.
As we will show, it is possible to make significant progress and 
 distinguish among many of these options even without 
new observations, using only archival data.

\section{Sample selection}

To define our sample of `blank field sources' 
we searched the {\it ROSAT} PSPC ``WGACAT95'' (or simply `WGACAT',
White Giommi \& Angelini  1995), a catalog containing a total of $\sim 62,000$ sources
  generated from {\em ROSAT} PSPC pointed observations
 using a sliding cell detect algorithm.
We used the following selection criteria:
\begin{enumerate}
\item bright X-ray sources, with $f_X >10^{-13}$ erg cm$^{-2}$ s$^{-1}$;
\item well detected, with a  signal to noise ratio greater than 10 and a 
quality flag\footnote{Since the sliding cell algorithm is sensitive to 
point sources, it can find spurious sources where extended emission 
is present.
WGACAT includes a quality flag, DQFLAG, that notes dubious detections 
based on a visual inspection of the fields.} greater than 5;
\item high Galactic latitude ($|b| > 20^{\circ}$);
\item not within $2^{\prime}$ of the target position 
(at which point the source density reaches the background level) 
to select only random, serendipitous sources;
\item location in the `inner circle' of the PSPC ($r<18^{\prime}$) to
have smaller positional error circles;
\item north of $\delta = -18^{\circ}$ in order to have  measurements 
in the Automated Plate-measuring Machine (APM) catalog of the POSS
 (McMahon \& Irwin 1992);
\item unidentified, with WGACAT class= 9999 and no SIMBAD or NASA
Extragalactic Database identification.
\end{enumerate}

The first six criteria selected 1624 sources ($\sim 3$\% of the total).
 Since these sources were 
selected purely on their X-ray properties, they form a well defined  
sample from which to study the incidence of minority X-ray populations.
Adding the requirement that a source has to be unidentified left 940 objects.
We than searched the APM catalog for 
 sources without any optical counterpart in the O filter (to O=21.5)
in a radius of $26^{\prime \prime}$, which corresponds to about 95\% 
confidence for X-ray sources within $18^{\prime}$ of the PSPC detector 
center (Boyle et al. 1995).

We note that with a B band (similar to the  Palomar O-band) source density of 
1000-2000 sources deg$^{-2}$ at $B \sim 21.5-22$ (e.g. Prandoni et al. 1999;
Huang et al. 2001) there is a $\sim 15-35$\% probability for each PSPC 
26$^{\prime \prime}$ error circle \footnote{$\sim 36-74$\% when using the 
39$^{\prime \prime}$ 3$\sigma$ radius error circle (see below).}
of containing a source by chance.
This implies that many other PSPC sources might be `blanks' if we had a smaller
uncertainty on the X-ray positions.
The {\em ROSAT} HRI, thanks to its smaller positional uncertainty, would
have been a better choice for the selection of blanks,
but no HRI catalogs were available when this project was initiated.

This gave 81 sources.
After visual inspection  10 of these sources proved to be false
due to limitations of the detection algorithm: there
 can be false detections near bright sources
or detections close to the inner circle boundary due 
to the spread of a bright source falling in the outer circle.
The final sample is thus composed of 71 sources.

However careful inspection showed that some of the 71 {\em ROSAT} fields 
had WGACAT coordinates not in agreement with 
other position estimates and in some cases the displacement was  larger 
than the $26^{\prime \prime}$ radius error circle used for the 
cross-correlation. Some of our blanks had  no optical 
counterparts because of the wrong X-ray positions.
This shift in WGACAT coordinates is related to an error in 
converting the header of the old format event files into RDF format; 
this error caused an offset in the source declination up to 1$^{\prime}$.
These problems were solved in the newly released Revision.2 {\em ROSAT} PSPC data, 
available in the HEASARC database.

\subsection{Revised Sample}
\label{blanks_revisedsample}

In order to quantify how many sources had WGACAT coordinates with an 
offset $\geq 26^{\prime \prime}$ we used the incomplete Galpipe
catalog (Fabbiano et al. in preparation) and 
found improved Revision.2 coordinates
for 54 of the 71 objects (76\% of the sample).
Out of these, 39 had a shift between WGACAT and Galpipe 
positions smaller than $26^{\prime \prime}$.
(Note that even if $\sim 30$\% of the sources in our sample 
have an offset $>26^{\prime \prime}$ 
only 431 observations out of 3644 ($\sim 12$\%) in WGACAT were 
affected by this problem.)
 Our method was highly efficient at selecting coordinate
errors, as well as true `blanks'.
Since no {\em ROSAT} PSPC catalog based on Rev.2 images was completed at that
time\footnote{A second version of the WGACAT based on rev. 2 data was 
released on May 20, 2000 (WGACAT00). 
On the same day an updated version of WGACAT95  was released.
The positional problem in WGACAT95 was solved thanks to our note to the authors
(Cagnoni, White, Giommi \& Angelini 1998, private communication).},
it was not possible to perform the whole selection using the 
correct coordinates.
We repeated the cross-correlations with the POSS using the Galpipe 
coordinates of the 54 WGACAT-Galpipe sources using 
a more conservative $39^{\prime \prime}$ radius error circle, corresponding
to $3 \sigma$ confidence (Boyle et al. 1995).
We have been left with the 16 \bfs \/ 
listed in Table~\ref{blanks_tab_16src}.
As a result of this change in definition, our sample is not
statistically complete.\\
Note also that clearly 
X-ray bright sources could still have high \fxv $> 10$ and not be 
included in our sample if their optical counterpart exceedes the O$>$21.5 
limit.

\section{{\em ROSAT} Data on Blank Field Sources}
\label{blanks_pspc}

We searched the {\em ROSAT} archive for all  PSPC and HRI 
observations of the final sample of the 16 `blanks'
 listed in Table~\ref{blanks_tab_16src}.
Besides the PSPC observations used for the selection, 
six `blanks' have other serendipitous 
detections in both the PSPC and HRI.
A list of all the {\em ROSAT}  observations is presented in
Table~\ref{blanks_pspc_hri}.\\
To determine the rough X-ray spectral properties of our blanks
we measured their X-ray hardness (HR=H/M) and softness (SR=S/M)
ratios based on the WGACAT95 count rates in the standard {\em ROSAT} PSPC bands: Soft, S (0.1-0.4~keV), Medium, M (0.4-0.86~keV), and Hard, H
(0.87-2~keV). We converted these  ratios into effective X-ray
spectral indices $\alpha_{S}$ (0.1--0.8 keV) and $\alpha_{H}$ (0.4--2.4 keV),
to correct for the variable Galactic line-of-sight absorption,
and the energy-dependent PSF, following Fiore et al. (1998).
Note that the effective spectral indices are not physical spectral
 slopes, but should be considered analogous to U-B and B-V colors, 
(see Fiore et al. 1998 for a detailed discussion).  Due to the low
signal-to-noise ratio of X-ray data, individual spectral indices
are not reliable but they are useful on a statistical basis.

\begin{figure}
\plottwo{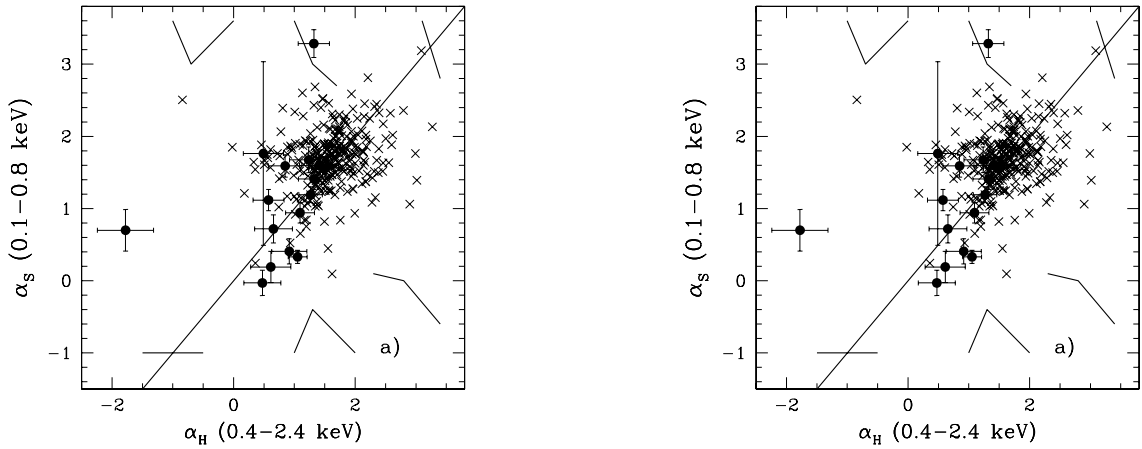}{icagnoni_f1a.eps}
\caption{\footnotesize{{\it ROSAT} PSPC effective spectral indices of \bfs \/ (filled circles) 
derived as in Fiore et al. (1998), compared to a reference sample of 
(a) radio-quiet quasars and to (b) radio-loud quasars from Fiore et al. 
(1998) (crosses). Three-point spectra illustrate the 
radically different spectral shapes in different parts of the diagrams.} \label{blanks_effalpha}}

\end{figure}

In Figure~\ref{blanks_effalpha} we compare the effective spectral indices 
of the blanks (filled  circles) with (a) those of a sample of 
radio-quiet quasars and (b) of a sample of radio-loud quasars from 
Fiore et al. (1998), 
as they derive the effective spectral  indices in the same way and 
using the same correction factors to the WGACAT hardness ratios.
Blanks tend to have harder spectra, both in the soft and hard PSPC 
bands, when compared to the radio-quiet quasars, while their distribution 
is closer to that of the radio-loud ones.
Could this be due to a radio-loud nature  for the \bfs ? This is not confirmed
by  radio data (\S~\ref{blanks_radio}).

We simulated the spectra of the main classes of X-ray sources to find 
their place in
the $\alpha_S-\alpha_H$ plane (Figure~\ref{blanks_effalpha}).
We modeled stars using a black body spectrum; for $kT \leq 70$ eV 
they have  steep PSPC spectra with 
 $\alpha_S \geq 2 $ and $\alpha_H>4$, and thus would fall outside 
of the region plotted in Figure~\ref{blanks_effalpha} 
in the upper right corner direction.
To simulate normal galaxies and local clusters of galaxies 
we used a Raymond-Smith thermal spectrum with the temperature ranging 
from 1 to 10 keV. Such sources lie in
a region with $-0.7 \leq \alpha _S \leq 0.2$ and $0 \leq \alpha _S \leq 0.7$.
For  $kT \sim 1$~keV local groups and clusters are thus expected to 
contaminate the  \bfs \/ falling below the diagonal line in 
Figure~\ref{blanks_effealpha}b.
Increasing the temperature 
would shift them down in the plot parallel to the diagonal line and out of the
region occupied by the blanks.
We also simulated Seyfert-2s with an absorbed power law model ($N_H =10^{21}$ 
cm$^{-2}$ and  $10^{22}$ cm$^{-2}$).
As absorption cuts-off their low energy spectra, Seyfert-2s 
lie in the  $\alpha_S < \alpha_H$ 
region of the  diagram: for $N_H =10^{21}$ cm$^{-2}$ 
  near the highest temperature local clusters 
(below the diagonal line
 at $\alpha _H \sim 0$); increasing the $N_H$  makes 
them shift out of the plotted 
region on a diagonal path.
The BL Lac spectral shape depends on the position of  the SED  peaks 
with respect to the X-ray band: we adopt  a simple power law model 
with $\alpha_S = \alpha_H$ (diagonal line in Figure~\ref{blanks_effalpha})
 as a good  representation 
(e.g. Maraschi et al. 1995; Urry et al. 1996).

\subsection{PSPC X-ray Spectra}
\label{blanks_spectra}

We extracted the spectra of the 16 blanks from the archival Revision.2 
{\it ROSAT} PSPC data at  HEASARC using XSELECT 2.0.
When the source was detected in more than one observation we 
combined the spectra.
We used as source extraction region a circle with a radius of 160 pixels 
($1^{\prime} 20^{\prime \prime}$) and as background an annular 
region centered on the object (radii $1.6 -3.2^{\prime}$) from which we 
removed  any obvious sources.  
 We fit the PSPC spectra of each source using
 XSPEC 11.0.1  with three different models:\\

(1) an \underline{absorbed power law} model of the form\\
\[ F(E)= k \times E^{-\Gamma} \exp{-[ \rm{N_H} {\sigma}_H ]}\]
where $F(E)$ is in photons cm$^{-2}$ s$^{-1}$ keV$^{-1}$,
 $\Gamma$ is the photon index, $k$ is the normalization at 1~keV 
and the (local) 
absorption is characterized by  a column density $N_H$ and an
absorption cross section $\sigma_{\rm{H}}$. 
We fitted the data leaving all the parameters free and
again fixing the absorbing column to the Galactic value 
from the Bell Labs H{\sc i} survey (Stark et al. 1992, found  with
{\em w3nh} at HEASARC).\\

\begin{figure}
\epsscale{1.0}
\plotone{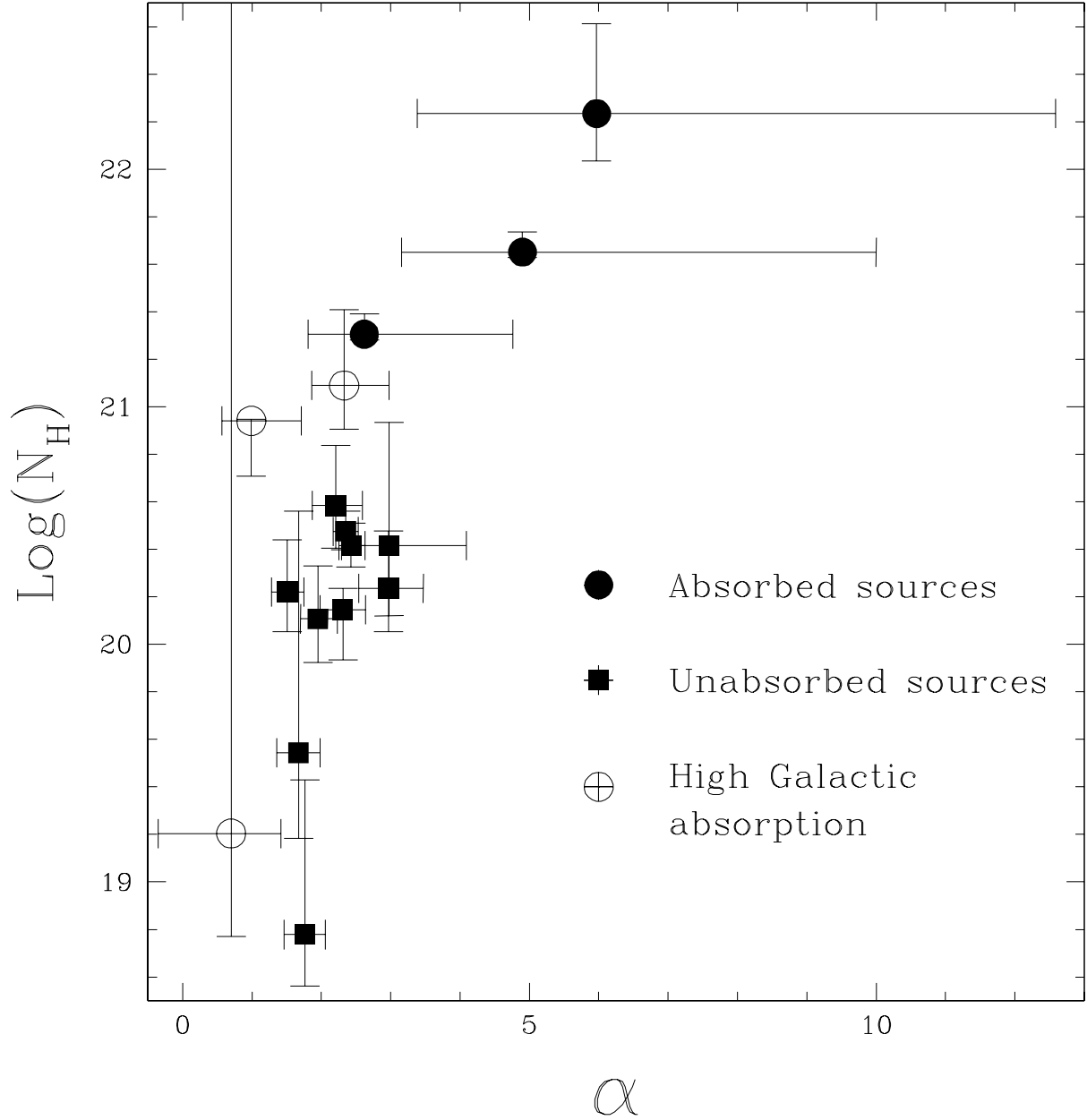}
\caption{\footnotesize{Absorption column versus energy spectral slope as 
derived from the fit with an absorbed power law model. 
Filled circles represent sources with
indication of heavy absorption in excess of the Galactic value; open circles
sources with high Galactic column density and  squares the sources with 
absorption consistent with the Galactic value.}\label{blanks_alpha_NH}}
\end{figure}

The fit values of $\Gamma$ and $N_H$ are plotted for each source in
Figure~\ref{blanks_alpha_NH}. As evident in Figure~\ref{blanks_alpha_NH} the power law fit gives a
$\geq 3 \sigma$ indication, for an absorbing column in excess of the 
Galactic value for three \bfs \/ 
(filled points in Figure~\ref{blanks_alpha_NH}).
It appears that, omitting three sources with high Galactic $N_H$ 
($> 4.26 \times 10^{20}$ cm$^{-2}$), 
the spectra fall into two groups:\\
\indent (a) - the \underline{`absorbed' sample}: 
sources with an indication of 
absorption in excess of the Galactic value ($\alpha_H \leq 0.62$ and 
$\alpha_S \leq 0.70$), including 
 three  sources 
(1WGA~J1243.6$+$3204; 1WGA~J1216.9$+$3743 and 1WGA~J0951.4$+$3916);\\
\indent (b) - the \underline{`unabsorbed' sample}: 
sources with absorption consistent
with the Galactic value ($\alpha_H > 0.62$ and 
$\alpha_S > 0.70$).\\

(2) an absorbed \underline {black body} model of the form
\[ F(E) =  k \times 8.0525  E^2 dE / ((kT)^4 (\exp(E/kT)-1))  \exp{-[\rm{N_H} {\sigma}_H ]}\]
where $k$ is the normalization and $kT$ is the temperature in keV.
1WGA~J1243.6$+$3204 and 1WGA~J1216.9$+$3743 still require high absorption, 
while  1WGA~J0951.4$+$3916 is consistent with the Galactic value.\\

\begin{figure}
\plottwo{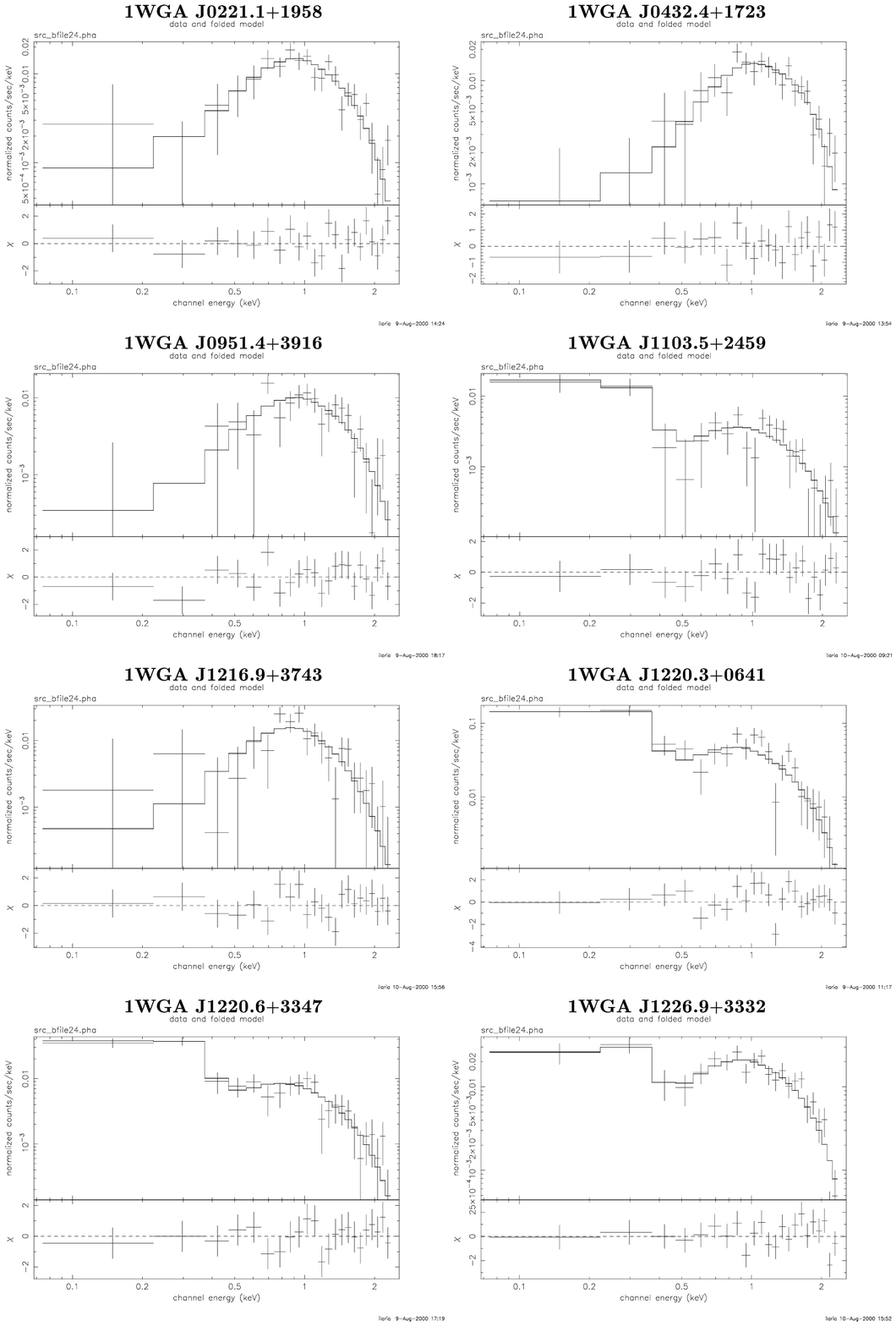}{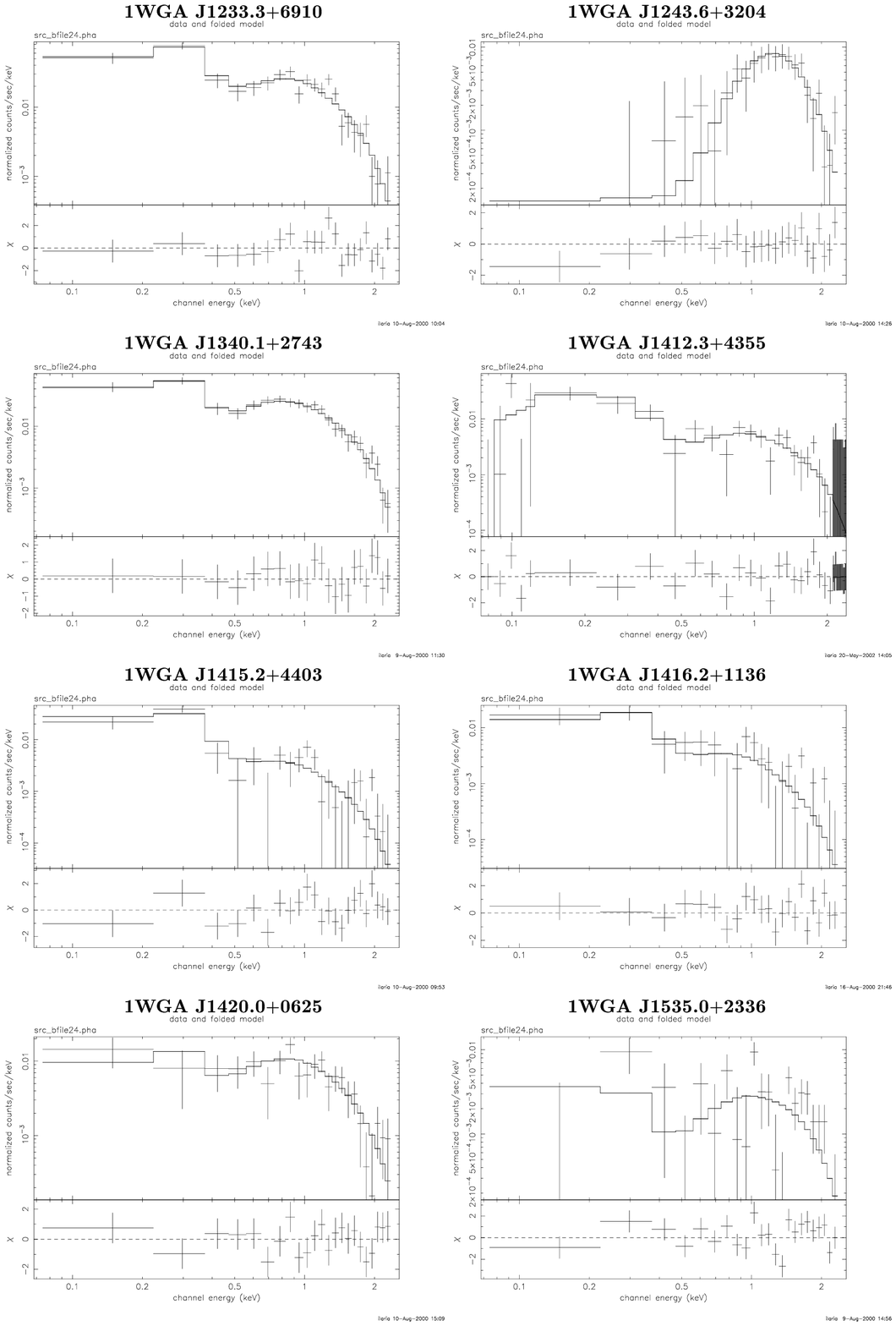}
\caption{\footnotesize{{\em ROSAT} PSPC energy spectra of the 16 \bfs \/ 
(upper panels); 
the solid lines represents the best fit with an absorbed power law model.
The lower panels show the residuals.}\label{blanks_singlespectra}}
\end{figure}

(3) an absorbed \underline{optically thin thermal plasma} (Raymond \& Smith, 1977) with absorption fixed  to the Galactic value, abundances fixed to 0.1 times the Solar value and redshift free to vary between 0 and 2.\\
The results of these fits are reported in Table~\ref{blanks_tab_fits}
 and the spectra of the blanks fitted with the absorbed power 
law model and their residuals are reported in 
Figure~\ref{blanks_singlespectra}. 

The spectra of the individual sources do not have enough  statistics 
 to discriminate between the competing models.\\

\subsubsection{Combined spectra}
\label{blanks_combinedspectra_section}

\begin{figure}
\plotone{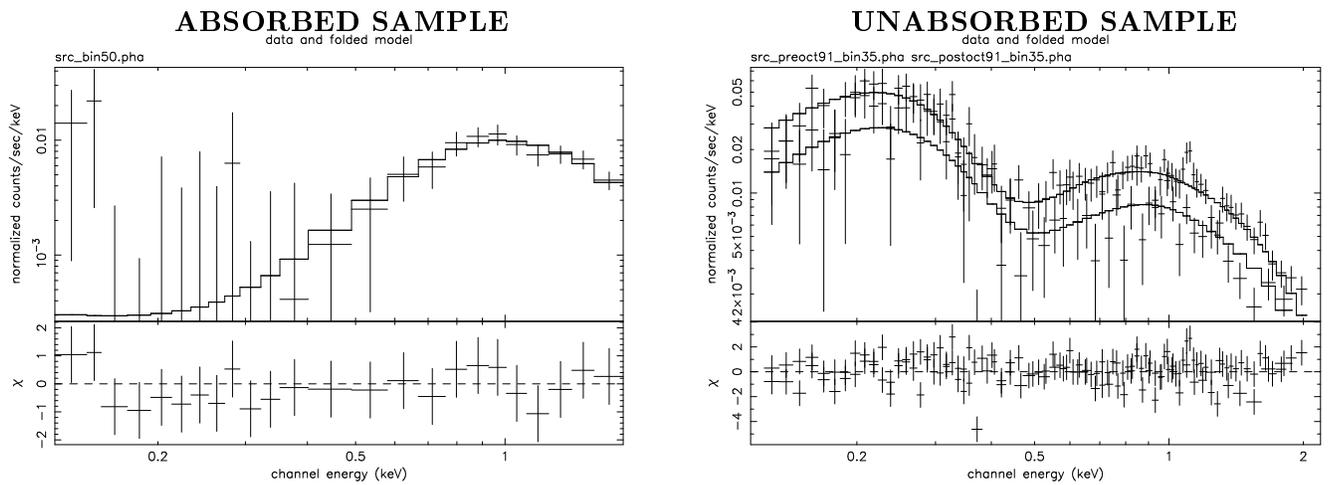}
\caption{\footnotesize{{\em ROSAT} PSPC combined energy spectra 
of the `absorbed' (left) and `unabsorbed' (right) samples and the 
residuals to the fit with an absorbed power law model with absorption fixed to 
the  mean Galactic value of the sample (bottom panels) (see text for details).}\label{blanks_combinedspectra}}
\end{figure}

In order to gain statistics,
we computed the average spectra of these two groups of
absorbed and unabsorbed sources (Figure~\ref{blanks_combinedspectra}). 
We sub-divided the unabsorbed sources into two
groups, i.e. the sources observed before and after the change in the PSPC
gain occurred on October 1991, computed two separate combined spectra
and then fitted the two spectra simultaneously. 
All the sources in the absorbed sample
were observed after October 1991.

By fitting these composite spectra with the models described in 
 \S~\ref{blanks_spectra}, 
we obtained the results summarized in Table~\ref{blanks_tab_fits}.
The best fit model for both  samples is an absorbed power law 
with photon index $\sim 2$.
The `absorbed' sample requires 
$N_H \sim 2 \times 10^{21}$ cm$^{-2}$, well
in excess of the Galactic value, while for the `unabsorbed' sample a 
Galactic column density is acceptable.
The Raymond-Smith model provides a good description of the data for both the
unabsorbed and absorbed (once the column density is left free to vary) sample.
The black body model is acceptable for the
`absorbed' sample, but it completely fails to describe the `unabsorbed' sources.
The large $\chi ^2$ values obtained for the `unabsorbed' sample suggests 
a more complex spectral shape, probably because different classes of sources 
are involved.

\subsection{X-ray fluxes}
\label{sec_fluxes}

We computed the absorbed X-ray fluxes in the 0.5-2.0 keV (the standard 
{\em ROSAT} band), 0.05-2.5 keV  (the WGACAT95 band) and 
0.3-3.5 keV ({\em Einstein} band) using the best fit absorbed power 
law model for each source (with the free absorption value, 
 except for 1WGA~J1103.5$+$2459, 
for which freeing the absorption leads to a
value $\sim 4$ time smaller than the Galactic value). 
The results are reported in Table~\ref{blanks_fx}.

For 4 sources 
(1WGA~J1416.2$+$1136, 1WGA~J1535.0$+$2336,
1WGA~J1103.5$+$2459 and 1WGA~1415.2$+$4403) the 0.3-3.5 keV flux we
 estimated is $\sim 40$\% lower than the WGACAT95 flux.
 Since $f_X(0.3-3.5 keV) <10^{-13}$ \cgs \/ for these sources,  their
$f_X/f_V$ is less  extreme and compatible with normal type~1 AGNs.
However this discrepancy is due to a steep  (for 1WGA~J1416.2$+$1136, 1WGA~J1103.5$+$2459, and 1WGA~J1415.2$+$4403) or a flat spectral slope (1WGA~J1535.0$+$2336)
compared to the average type~1 AGN spectrum for which the WGACAT 
count rate-to-flux conversion factor was estimated.
Instead five sources 
(1WGA~0432.4$+$1723, 1WGA~1220.3$+$0641, 1WGA~1226.9$+$3332, 
 1WGA~1233.3$+$6910 and 1WGA~J1340.1$+$2743) have 
larger $f_X$ than in the WGACAT.
With $f_X(0.3 -3.5 {\rm keV}) > 4 \times 10^{-13}$ \cgs  and lack of a POSS counterpart,  their 
\fxv \/ is not even compatible with extreme BL Lacertae objects.

\subsection{Variability}
\label{blanks_variability}

We extracted the lightcurves for sources and backgrounds (using the same
regions chosen for the spectral analysis) within each PSPC 
and HRI observation.
We binned the lightcurves with bin sizes ranging from 1000~s to 5000~s
and calculated the maximum variation factor of the count rate
considering the points with errorbars smaller than 
40\%.
The average maximum variability factor is $1.60 \pm 0.71$, 
consistent with the sources being constant.
The variability factor was computed using all
the available lightcurves spanning a minimum time scale of 1ks 
to a maximum time scale of $\sim 2$ years.
No strong variability is seen.
Neither do we see variability when comparing multiple PSPC and/or HRI
observations, for the 10 sources with multiple observations: the
count rates are consistent, 
within the large errors due to spectral uncertainties.\\
We also searched the RASS and three blanks
were detected in 1990 (Table~\ref{tab_rass}); all of them have RASS count rate consistent within 
1$\sigma$ with the pointed observations except for 
1WGA~J1233.3$+$6910,
 whose measured count rates can be reconciled within 2$\sigma$ 
(a factor 1.6 variability; Table~\ref{tab_rass}).
An extremely variable nature, transient or burst-like,  is
excluded for all the 16 blanks.

\subsection{ROSAT HRI: extent}
\label{blanks_hri}

7 sources have one or more {\em ROSAT} HRI observations (see 
Table~\ref{blanks_pspc_hri}).
Only one source (1WGA~J1216.9+3743) is not visible in the HRI images.
The detected sources look pointlike.  1WGA~J1226.9+3332 
(\S ~\ref{blanks_thecluster})  has an extension at the limit of the HRI 
PSF ($\sim 5^{\prime \prime}$) and  looks extended in a {\em Chandra}
follow-up observation (Cagnoni et al. 2001b).
 1WGA~J1233.3+6910, instead,  falls on the detector's edge
making any estimate of count rate, position and extent unreliable.
The HRI positions and count rates are reported in 
Table~\ref{blanks_tab_16src}.


\section{ASCA}
\label{blanks_asca}

We searched the ASCA public archive for serendipitous observations of the 
16 blanks and we found that 4 of them (1WGA~J1226.9+3332, 1WGA~J1535.0+2336, 
1WGA~J1415.2+4403 and  1WGA~J1220.3+0641) fall in the ASCA GIS
field of view (see the log in Table~\ref{blanks_tab_asca}.
All the sources but 1WGA~J1415.2+4403 are detected, but 1WGA~J1226.9+3332
and  1WGA~J1535.0+2336 are too faint to derive spectra.
For 1WGA~J1220.3+0641, instead, {\em ASCA} data  provide valuable broad X-ray spectral information.
The results are discussed in the sections dedicated to each source (\S ~\ref{blanks_individual}).

\section{Radio}
\label{blanks_radio}

To check on the radio emission of the blanks
we searched for radio counterparts in the
NRAO VLA Sky Survey (NVSS, Condon et al. 1998)
and in the FIRST survey (Becker, White \& Helfand 1995) 
within the $39^{\prime \prime}$ X-ray error circle. 
Both surveys were conducted at 1.4 GHz.
The NVSS has a limiting flux of $\sim 2.5$~mJy 
and a position uncertainty of few arcseconds.
The FIRST 
ongoing survey reaches $\sim 1$ mJy with a position
accuracy $<1^{\prime \prime}$. 
We found 4 NVSS counterparts for
the blanks, 2 of which detected also in the FIRST. 
The remaining 2 NVSS sources are
in an area not yet covered by  FIRST (see 
Table~\ref{blanks_tab_radio}).
We computed the broad band spectral indices  $\alpha_{ro}$ and $\alpha_{ox}$ (similar to \fxv ) for the 4 sources with dario data
using the NVSS radio flux,
the X-ray flux at 1 keV (see Table~\ref{blanks_tab_fits}) 
and a O magnitude limit of 21.5 (considered to be equal to the V
 magnitude).\footnote{We used NVSS fluxes instead of FIRST fluxes because the
latter ones can be  underestimated because
of the high spatial resolution of the survey.
Even in the NVSS,
 some of the flux coming from extended sources could be either resolved
out or split into two or more components, leading to a systematic underestimate
of the real flux densities of such sources (Condon et al. 1998).}

For 1WGA~J1226.9$+$3332, identified as a z=0.89 cluster of galaxies
(see \S~\ref{blanks_thecluster}) we measured a R band magnitude of 20.4
and we converted it into V magnitude of 22.2 
using  (V-R)=1.83, typical for an
elliptical galaxy at z=0.89 (Coleman, Wu \& Weedman 1980), obtaining
$\alpha_{ro}= 0.60 $ and $\alpha_{ox}=0.68$.
For 1WGA~J1340.1$+$2743, a BL Lac (see \S ~\ref{1WGAJ134012743})
 we used $R \sim 22$
 (from Lamer, Brunner \& Staubert 1997) and converted it using the mean
$V-R \sim 0.65$ found for BL Lacs (Moles et al. 1985) and we
obtained $\alpha_{ro}= 0.63 $ 
and $\alpha_{ox}=0.61$.
Both sources are radio-loud with $f_{({\rm 1.4 GHz})}/f_V \sim 1000$.\\
For the remaining two blanks with a radio counterpart we  obtained 
values of $\alpha_{ro} \geq 0.5$ and $\alpha_{ox} \leq 0.7$.

We allso translated the \fxv \/ ratio into $\alpha_{ox}$ 
for the remaining sources.\\
Since BL Lacs are the known class of X-ray emitting sources with
higher $f_X/f_V$, we compared the $\alpha_{ro}$, $\alpha_{ox}$
with those of X-ray selected BL Lacs 
(e.g. Caccianiga et al. 1999).
Figure~\ref{blanks_caccia} shows that 
the blanks with a radio counterpart occupy an extreme region of the
$\alpha_{ro}$-$\alpha_{ox}$ plane, where just few BL Lac objects
fall.
Deeper radio observations would be of great value.

\section{Discussion and notes on individual sources}
\label{blanks_individual}

In this section  discuss all the possibilities 
for the blanks nature  presented in \S ~\ref{blanks_nature}.
This section also includes all the available information on the sources 
at the time of writing.
While all of them were unidentified when the 
sample was first selected, some were studied
(and in some cases identified) by other groups while we were progressing
 with the investigation.
A schematic summary of the identified sources and a possible classification 
for the still unidentified ones is presented in Table~\ref{blanks_tab_ids}.

\subsection{BL Lacertae objects}

\underline{1WGA~J1340.1$+$2743}\\		
\label{1WGAJ134012743}
1WGA~J1340.1$+$2743 was observed 7 times with the PSPC, only once 
in the inner circle.
We extracted a combined image and spectrum.
This source is close ($\sim 7^{\prime}$)  to the bright Seyfert 2
galaxy  CC~B00 (1RXS~J134021.4+274100) which has $\sim 2$
times 1WGA~J1340.1$+$2743 count rate.
Due to slight displacements in the source positions as detected in
different images and to the broader PSF in the outer regions of the
detector, 1WGA~J1340.1$+$2743
is contaminated by the nearby bright 
source making the combined  image not suitable for variability
or flux estimates.

1WGA~J1340.1$+$2743 was identified as a BL Lac in 
 Lamer, Brunner \& Staubert (1997) because of lack of spectral
features in its optical spectrum.
 Lamer et al. find an X-ray flux at 1 keV of 
$2.06 \times 10^{-14}$ erg cm$^{-2}$ s$^{-1}$.
in a {\em ROSAT} observation 
 during which the source falls in the PSPC outer rim.
This value is 50\% smaller than the 
$3.07 \times 10^{-14}$ erg cm$^{-2}$  s$^{-1}$ we find for the
 detection in the inner PSPC circle.
1WGA~J1340.1$+$2743 has a radio counterpart  $\sim 10^{\prime \prime}$
from the PSPC position (see \S ~\ref{blanks_radio} and 
Table~\ref{blanks_tab_radio}). However  the $\alpha _{ro} = 0.62$ and $\alpha
_{ox} = 0.61$ (see \S ~\ref{blanks_radio} for details)
 and the $f_X/f_V \sim 300$ are  not compatible with the values of 
`normal' BL Lacs.
An extreme BL Lac (e.g. Costamante et al. 2001)
with the synchrotron  peak lying,  at energies higher than the 
{\em ROSAT} band,  could fit this source.
The X-ray spectral slope ($\Gamma = 2.35 \pm 0.18$) is also similar to those 
found with {\em Beppo-SAX} by Costamante et al. (2001) for extreme BL Lacs.\\
Another possibility is that  1WGA~J1340.1$+$2743 is an ``average'' X-ray 
selected BL Lac, that was undergoing a large flare 
during the WGACAT95 detection:
a factor of 10-30 variation has been observed in X-ray selected BL Lacs 
(e.g. MRK~501, Pian et al. 1998) and a 10 times smaller ``quiescent flux''
would drastically lower the  $f_X/f_V$ down to $\sim 20$, a normal value
for BL Lacs.
This hypothesis, however, is not confirmed by the PSPC lightcurve, 
where no dramatic changes in the X-ray flux
are visible other than a factor of 2 brighter with respect to the results by
 Lamer, Brunner \& Staubert (1997).

\subsection{Isolated neutron stars}
\label{blanks_ions}

As mentioned in \S \ref{sec_ins}, 
the spectrum expected from INSs  is thermal,  with 
typical temperatures $T \sim 100$~eV (but models with harder spectra have 
also been proposed, e.g. Zampieri et al. 1995).
The flat spectra of blanks suggests that statistically it is 
unlike that they are INSs, even though INSs could hide
among the softest sources in our sample.
With a well-defined statistical sample it will be possible
to set limits on the surface and space density of INSs.

\subsection{The AGNs and AGNs candidates}
\label{blanks_individual_agns}
\underline{1WGA~J1220.3$+$0641}\\		
\label{1WGAJ122030641}
1WGA~J1220.3$+$0641 was detected in the RASS and
it is  listed in 
the Bright Source Catalog (BRASS, Voges et al. 1999) with a broad band
 $CR=0.079 \pm 0.017$ counts s$^{-1}$.
Using the whole 0.07- 2.4 keV  {\rm ROSAT} band we derive a $CR = 0.098 \pm 0.006$ 
counts s$^{-1}$ consistent with the RASS detection. 
Its position in the BRASS catalog is 
14$^{\prime \prime}$ from the Galpipe position and  10$^{\prime
\prime}$  from the HRI position .
The HRI and Galpipe positions for this source are only 
 $4^{\prime \prime}$  apart.\\
{\em ASCA} found 1WGA~J1220.3$+$0641  as an unidentified 
serendipitous source in 1995 in the {\em ASCA} GIS with a 1-10 keV
 count rate of $0.0142 \pm 0.0009$ (George et al. 2000).
This source is also included in the ASCA Medium Sensitivity Survey (AMSS)
(Ueda et al. 2001) with the name 1AXG~J122017$+$0641.
The {\em ASCA} archival lightcurves do not
show any strong variability (Figure~\ref{blanks_asca_lc}).
In the AMSS the source has a count rate of 0.0214 counts s$^{-1}$ in the 0.7-7.0 keV band,
corresponding to a flux, corrected for the Galactic absorption,
 of $1.4 \times 10^{-12}$ \cgs , with most of the flux 
lying in the 2-10 keV band.
An absorbed power law model with absorption fixed to the Galactic value 
($1.56 \times 10^{20}$ cm$^{-2}$)
well describes the source; while freeing the absorption does 
not improve the fit.
{\em ASCA} GIS2 and GIS3 data  give values of
$\Gamma = 1.31 \pm 0.58$
and normalization of $4.16 \times 10^{-4}$ photons cm$^{-2}$ s$^{-1}$
 keV$^{-1}$ at 1 keV, with a reduced $\chi ^2$ of 0.79 (33 d.o.f.).
The simultaneous fit of {\em ROSAT} PSPC  and 
{\em ASCA} GIS2  and GIS3  spectra gives a best fit photon index of
1.18 and a normalization of $3.38 \times 10^{-4}$ 
photons cm$^{-2}$ s$^{-1}$ keV$^{-1}$
at 1 keV, with a reduced $\chi ^2$ of 2.608 (56 d.o.f.).
The data and the unfolded spectra 
fitted with an absorbed power law model are presented in
 Figure~\ref{blanks_asca_sp}.\\
If the absorbed power law is a  correct model, 
1WGA~J1220.3$+$0641 looks like a normal type~1 AGN.

The POSS shows the presence of a O=21.81 and E=18.58 red
source within the 1WGA~J1220.3$+$0641 error circle.
A local source with O-E=3.23 would be strongly obscured 
 ($A_V \sim 6$)  if an intrinsic quasar spectral shape
(e.g. Francis et al. 1991) 
is adopted (e.g. Figure~7 in Risaliti et al. 2001).



\underline{1WGA~1412.3$+$4355}\\			
\label{1WGA141234355}	
This source has been  classified as an AGN at $z \sim 0.59$
in the RIXOS survey (Mason et al. 2000) on the basis of one broad optical
emission line identified 
with  Mg~{\sc ii}. The RIXOS (0.5 - 2.0 keV) count rate is
$0.0055 \pm 0.006$ counts s$^{-1}$, compatible with our measurement.
The X-ray photon spectral slope we find is $2.02 \pm {0.15}$
(c.f. $2.1 \pm 0.1$ in the RIXOS)
 consistent with a normal type~1 AGN
and the spectrum does not show  signs of an absorbing column in excess of the low Galactic value
($1.17 \times 10^{20}$ cm$^{-2}$).
1WGA~1415.2$+$4403 is not variable in the PSPC observation, but there are no 
other observations for this source.\\
The POSS shows a E=18.73 source within the error circle, but no 
objects are detected in the O-band; assuming a limit of O$>$21.5, we obtain a  color of
O-E$>2.77$, corresponding to   $A_V > 4$ at $z=0.59$ for a quasar.


\underline{1WGA~1415.2$+$4403}\\			
\label{1WGA141524403}
1WGA~1415.2$+$4403 is also included in the RIXOS survey (Mason et al. 2000)
and classified as a $z=0.562$ AGN based on 3 optical lines.
The RIXOS (0.5 - 2.0 keV) count rate is 
$0.0033 \pm 0.0005$ counts s$^{-1}$, comparable to  our count rate 
($0.0030 \pm 0.0006$ counts s$^{-1}$).
The photon spectral slope is $2.70^{+0.22}_{-0.20}$
($2.3 \pm 0.1$ in the RIXOS) steeper than typical type~1 AGNs.
There are no signs of absorption in the X-ray spectrum.
This source is not variable in the PSPC observation.\\
1WGA~1415.2$+$4403 was serendipitously observed by the {\em ASCA}
satellite in 1996 (Table~\ref{blanks_tab_asca}) 
but the source was not detected.
The $3\sigma$ upper limit on the GIS2+GIS3 count rate in a 
$r=6^{\prime \prime}$ circle is $0.00155$ counts s$^{-1}$, consistent 
with the count rate predicted by PIMMS by extrapolating the
{\em ROSAT} PSPC count rate using the best fit power law with
 Galactic absorption (values  taken from
Table~\ref{blanks_tab_16src} and Table~\ref{blanks_tab_fits}).\\
No sources are found in the POSS O and E bands within the X-ray error 
circle.
We note that the WGACAT95 flux was $\sim 50$\% overestimated (see \S \ref{sec_fluxes}
This source is, as a consequence, one of the least extreme 
in our sample; 
the $f_X/f_V$ is $\sim 5$ and $f_X/f_R$ (Hornschemeier et al. 2001) $\sim 1$
are both consistent with the tail of normal type~1 AGNs.

\underline{1WGA~J1416.2$+$1136}\\		
\label{1WGAJ141621136}	
This source was found in the Cambridge ROSAT Serendipitous
Survey  (CRSS,
Boyle, Wilkes \& Elvis 1997) with a (0.5-2.0~keV) X-ray flux of
 $5.2 \times 10^{-14}$~erg cm$^{-2}$~s$^{-1}$\footnote{Using a constant 
conversion factor of 1 count s$^{-1}$ = $1.2
\times 10^{-11}$~erg~cm$^{-2}$~s$^{-1}$}  implying a count
rate of $0.433 \times 10^{-2}$ counts s$^{-1}$, larger than
 our count rate of $(0.31 \pm 0.08) \times 10^{-2}$ counts s$^{-1}$. 
In the {\it ROSAT} HRI the count rate is 
consistent with PIMMS estimates  based on the PSPC count rate suggesting a
constant source over a 4-year period.
Also for this source the WGACAT was overestimated ($\sim 60$\%) 
and the source is thus less extreme than previously thought.
The X-ray photon spectral index is $2.68^{+0.27}_{-0.24}$,
somewhat steep  for  a normal type~1 AGN.\\
On the POSS there is a O=22.22 and  E=19.50 source 
$12.8^{\prime \prime}$ from the HRI position.
The redness of the source O-E=2.72 implies $A_V \sim 5$ for a typical AGN spectrum (Francis et al. 1991), corresponding to
$N_H \sim 10^{22}$ cm$^{-2}$.
The {\em ROSAT} spectrum, however, is consistent with
Galactic absorption.




\underline{1WGA~J1535.0$+$2336}\\		
\label{1WGAJ153502336}	
This source was detected by the {\it Einstein}  IPC as an ultra-soft source
(Thompson, Shelton, \&  Arning 1998)  with a count rate of $(3.64 \pm
0.96) \times 10^{-3}$ counts s$^{-1}$, 
in good agreement with the PSPC
count rate using our best fit values for an absorbed power law model
 (Table~\ref{blanks_tab_fits}).
The {\it Einstein} and Galpipe positions are 17$^{\prime \prime}$ apart
and thus consistent.
The X-ray spectral slope ($\Gamma = 0.70^{+0.72}_{-1.05}$) 
is consistent within the large errorbars with a type~1 AGN.
1WGA~J1535.0$+$2336 was detected in 1993 by the {\em ASCA} GIS
(Table~\ref{blanks_tab_asca}) with a full band GIS2$+$GIS3 count rate of 
$\sim (8 \pm 3) \times 10^{-4}$ counts s$^{-1}$ in a $r=2.25^{\prime}$
radius circle. An extrapolation of the {\em ROSAT} PSPC count rate into the 
{\em ASCA} band using the flat best fit spectral slopes  
($\Gamma =0.7$  for a power law model with
free absorption)
overestimates the {\em ASCA} count rate by about an order of magnitude.
A steeper value, $\Gamma =2.2$, which still lies  within the $2\sigma$ 
uncertainty on the {\em ROSAT} fit, is needed to achieve consistency with {\em ASCA} and with the {\em EInstein} ultra-soft ($\Gamma > 2$) nature.
This is a typical value for type~1 AGNs.
A $\sim 2$ days timescale variability (the time distance between 
the {\em ROSAT} and {\em ASCA} observations) could also reconcile
the measured count rates.

1WGA~J1535.0$+$2336 has an extremely flat spectral slope
 and the AGN-like count rate to flux conversion
used in the WGACAT95 gives a  $\sim 90$\% overestimated flux value.\\
A O=21.79 and E=19.61 source is visible in the Palomar within 
the 1WGA~J1535.0$+$2336 error circle.
O-E=2.18 implies $A_V \sim 4$ for a local source 
and $N_H \sim 8 \times 10^{21}$ cm$^{-2}$ (assuming Galactic dust to gas
ratio), while the X-ray fit does not show the need for extra absorption.


\subsubsection{Discussion on the blank field AGNs}
\label{blanks_agns}

When we started this project we expected to find examples of
peculiar  AGNs such as absorbed ones (QSO-2s or LMSy2s) or low efficiency radiators
 (\S ~\ref{blanks_nature}).
Indeed two sources,  1WGA~J1412.3$+$4355 and 1WGA~J1415.2$+$4403
were spectroscopically identified in the RIXOS survey (Mason et al. 2000)
as AGNs and while the latter could be a normal type~1 AGN 
(see \S ~\ref{1WGA141524403}), the first one has an unexpected Palomar
 counterpart (see \S ~\ref{1WGA141234355}).
Another 5 sources (1WGA~J1220.3$+$0641, 1WGA~J1416.2$+$1136, 
1WGA~J1535.0$+$2336, 1WGA~J1233.3$+$6910 and 1WGA~J0951.4$+$3916,
see Table~\ref{blanks_tab_ids}) are likely to be AGNs.
1WGA~J0951.4$+$3916 stands out  because it
 shows signs of absorption in the X-ray band 
($N_H \sim 2^{+3}_{-1} \times 10^{21}$ cm$^{-2}$, 
Table~\ref{blanks_tab_fits}) and has a red source in the POSS
E-band  (E=19.1).

The remaining  4 AGN-like sources  seem to share the same peculiar combination
of characteristics: they have 
 X-ray spectra typical of  type~1 AGNs ($\Gamma \sim 2$  and 
absorbing column densities consistent with the Galactic values (except for
 1WGA~J0951.4$+$3916 which shows an X-ray derived $N_H \sim 2 \times 10^{21}$ cm$^{-2}$) 
and they 
have  red (O-E$>2$) counterparts in the POSS E-band
at $E<19.5$ (Table~\ref{blanks_tab_ids}) (with the exception of 
1WGA~J1415.2$+$4403).
For a local AGN type~1, $O-E>2$ implies $A_V > 4$ (e.g. Risaliti et al. 2001) 
and thus strong obscuration ($N_H > 2 \times 10^{21}$ cm$^{-2}$ 
for a dust to gas ratio typical of the ISM in our Galaxy),
in contrast to the results of the X-ray data.\\
Since the host galaxy is not visible, we can assume $z \geq 0.2$
for the AGN blanks; this implies $L_{X {\rm (0.3-3.5 keV)}} > 
7.5 \times 10^{42}$ erg s$^{-1}$ or a bolometric X-ray 
luminosity $L_X \sim 3.7 \times 10^{44}$ erg s$^{-1}$ for a type~1 AGN
(and a total mass $> 3 \times  10^6 M_{\odot}$)
which is too bright for the galaxy emission to be important.
We propose here  some possibilities to reconcile X-ray and POSS data.

\underline{High redshift}\\
As is clear from Figure~7 of Risaliti et al. (2001),
at $z \geq 3.5$ AGNs have $(O-E)>2$,
 even for the low Galactic $A_V$.
The X-ray luminosity of the blanks if they were at $z \sim 4$ would 
be $\geq 10^{47}$ erg s$^{-1}$, consistent only with a quasar.
This option would be excluded for the two sources spectroscopically identified in 
the RIXOS (1WGA~J1412.3$+$4355
and 1WGA~1415.2$+$4403), which have $z \sim 0.5 -0.6$ (Mason et al. 2000), but
While for one of these (1WGA~1415.2$+$4403) the 3 observed optical lines
 uniquely determine the redshift, for 1WGA~J1412.3$+$4355 the redshift 
is uncertain.
In fact only one optical line at $\sim 2800$ \AA \/ was observed and identified as
Mg~{\sc ii}; if the observed line were instead the Ly${\alpha}$ line
 (1216 \AA \/ rest frame), the redshift of the source would be $2.66$,
almost in agreement with the Palomar color of the counterpart.

\underline{High dust to gas ratio}\\
Another way to reconcile the lack of absorption in the X-rays (mainly due 
to neutral hydrogen) with the obscuration in the optical band 
(due to dust) suggested by the red POSS counterparts, 
is to assume an higher than Galactic dust to gas ratio.
We derived the $A_V$ for the 4 sources with a counterpart in the 
POSS E band using the colors in Table~\ref{blanks_tab_16src} and 
Figure~7b of Risaliti et al. (2001) assuming a  
local redshift, except for the two RIXOS AGNs with measured redshifts.
We then derived the $A_V/N_H$ 
 using the Galactic $N_H$ for each source (Stark et al. 1992) 
and compared it to the extrapolation of the expected variation 
of the  $\rm A_V/N_H$ ratio as a function of the dust-to-gas ratio
presented in Maiolino et al. (2001a) (their Figure~2).
We derive dust to gas ratios $\sim 40-60$ times the Galactic value
when  Solar metallicity is assumed.
Observational evidence however shows a tendency of AGNs to have 
lower than Galactic dust to gas ratios (e.g. Maccacaro, Perola \& Elvis 1982; 
Risaliti et al. 2001; 
Maiolino et al. 2001b).
Furthermore such  high  dust to gas ratio is physically  
difficult to justify: in fact, assuming a Galactic dust to gas ratio
$\sim 1$\%  ($M_{\rm gas Gal}=8 \times 10^9 M_{\odot}$, 
Zombeck 1990 and $M_{\rm dust} \sim 10^7 M_{\odot}$ for spiral galaxies,
Block et al. 1994) the dust to gas ratio needed to reconcile Palomar and X-ray 
absorptions implies $\sim 40-60$\% of dust to gas ratio, a value exceeding 
the $\sim 37$\% of giant stars, the most efficient sources of dust formation 
in the Universe (Whittet 1992).\\
Maiolino et al. (2001b) find an higher than Galactic 
 ratio for low luminosity objects ($L_X \sim 10^{41}$ erg s$^{-1}$)
and suggest that these  may be physically different from
the more luminous ``classical'' AGNs.
We explore this option later in this section.

\underline{A different grain size distribution}\\
A possibility that cannot be excluded  is a modified dust grain 
size distribution  with respect to the Galactic diffuse ISM
(Mathis, Rumpl \& Nordsieck 1977).
A strong dominance of small grains would increase the optical extinction
without significant modifications of the X-ray absorption.

\underline{A dusty warm absorber}\\
AGN with a lack of X-ray absorption and the presence of
optical obscuration have been detected previously 
(e.g., Brandt, Fabian \& Pounds 1996; Komossa \& Fink 1997). 
The solution favored for these objects 
is the presence of  dusty ionized
absorbers.
In this model the dust 
is located within the ionized material and no X-ray cold absorption is
detected.  
The presence of dust within the warm absorber is expected to 
flatten the X-ray spectrum (Komossa \& Bade 1998), but we note that 
the slope range of the blanks AGNs(candidates) 
($\Gamma = 0.70 - 2.97$) is consistent with 
the sources for which a dusty warm absorber was proposed
(e.g. IRAS 13349$+$2438: $\Gamma =2.81$, Brandt et al. 1996;
NGC 3227: $\Gamma = 1.19$, Komossa \& Bade 1998;
NGC 3786: $\Gamma =1.0$, Komossa \& Fink 1997).

\underline{Missing or shifted BBB}\\
\label{sec_adaf}
Galaxies with bright X-ray emission ($L_X \sim 10^{42} - 10^{43}$ erg s$^{-1}$)
but weak or absent AGN features in the optical band have been known to 
exist since {\em Einstein} observations (Elvis et al. 1981; Tananbaum et al. 1997), and are now being found in large numbers  
(e.g. Della Ceca et al. 2001; Hornschemeier et al. 2001, Fiore et al. 2001; Comastri et al. 2002).
 A way to explain the faintness of the optical counterparts for these 
sources and for the five AGN-like blanks is to assume that the AGN 
is intrinsically  optically faint.\\

The intrinsic faintness could be related to a highly suppressed BBB emission.
Such suppression happens in low radiative efficiency accretion 
(\S ~\ref{blanks_nature}).\\ 
To check if this is consistent with the extreme properties we observe in our 
blanks we derived the V magnitude and the $\alpha_{ro}$ and  $\alpha_{ox}$
from  the spectral fit to Sgr~A* 
(Figure~1 of Narayan et al. 1998), assuming a distance of $\sim 10$ kpc.\footnote{We get $f_{({\rm 1.4 GHz})} = 580$ mJy,
V=17.22 and $f_{({\rm 1 keV})} = 7.75 \times 10^{-4}$ photons 
cm$^{-2}$ s$^{-1}$ keV$^{-1}$, while 
the 0.3-3.5 keV flux is $\sim 10^{-11}$ \cgs \/ 
($L_{({\rm 0.8-2.5 keV})} = 1.55 \times 10^{34}$ erg s$^{-1}$ 
Narayan et al. 1998)}.
We derive $\alpha_{ro}\sim 0.61$,  
$\alpha_{ox} \sim 1.12$ and $f_X/f_V \sim 18$, which are compatible
with \bfs .
These values are only indicative as could readily be made more extreme 
by changing the parameters of the model.
If blanks are such systems, the low accretion rate, 
coupled with the observed  luminosity 
($L_X \geq 3.7 \times 10^{44}$ erg s$^{-1}$ at $z\geq 0.2$) 
implies  a large mass ($M \geq 3 \times 10^9 M_{\odot}$).\\

The BBB emission may appear to be suppressed if its peak is 
shifted out of the optical band, as for low mass black holes. 
High energy BBB have been already observed 
(e.g. in the  NLSy RE~J1034$+$396,  Puchnarewicz et al.  1995; 2001).
Even though we cannot exclude such a possibility,
we note that this option does not hold for the X-ray bright, 
optically dull galaxies (e.g. Comastri et al 2002):
even if shifted, the BBB emission would still provide
the photons necessary to ionize hydrogen and 
show characteristic AGN emission lines 
which are not detected.
Furthermore, if the most extreme  blanks were nonvariable NLSy, 
an higher $f_X/f_V$ would imply higher disk emission 
and thus an unlikely higher accretion rates (NLSy are already 
thought to accrete
close to the Eddington limit, e.g. Laor et al. 1997, Leighly 1999, 
Turner et al. 1999 and references therein).

\underline{Extremely variable AGNs}\\
None of the sources seems to be extremely variable within each PSPC or between different  
observations, but we cannot a-priori exclude a high X-ray state during the 
{\em ROSAT} observations.

\subsection{Clusters of galaxies}
\label{blanks_individual_clusters}


\underline{1WGA~J1226.9$+$3332}\\
\label{blanks_thecluster}
This source is listed in Radecke (1997).
Using the same PSPC data, he found a count rate of 
$2.20 \times 10^{-2}$ counts s$^{-1}$ slightly smaller than ours,
$(2.66 \pm 0.16) \times 10^{-2}$ counts s$^{-1}$.
Short term variability is not visible between the  PSPC observations taken
two weeks apart.
In the HRI image the source is faint with a  count rate of 
$(8.7 \pm 1.5)  \times 10^{-3}$ counts~s$^{-1}$, in agreement with
PIMMS predictions based on the PSPC observations, indicating a
non-variable source over a 4-year period.
The HRI image is not conclusive for extension information.
We obtained a 10~ks  Cycle~1
{\em Chandra} observation for 1WGA~1226.9+3332\/ (Cagnoni et al. 2001a)
confirming the high z (z=0.89) cluster nature of this blank (see also Ebeling et al. 2001).
1WGA~J1226.9$+$3332 was also serendipitously detected 
in 1998 and in 2000 by the {\em ASCA} GIS.
The source falls at the border of the useful area of the GIS2 and GIS3 
detector and its  count rate
(GIS2$+$GIS3 full band count rate = $0.0038 \pm 0.0005$ count s$^{-1}$) 
should be considered a lower limit, but it probes to be close 
to the expected one.
(based on a PIMMS simulation with a Raymond-Smith model with $kT=10$ keV
and absorption fixed to the Galactic value).

\underline{1WGA~J0221.1$+$1958}\\		
\label{1WGAJ022111958}		
1WGA~J0221.1$+$1958 was identified in the SHARC survey (Romer et al. 2000)
as a $z=0.45$ cluster of galaxies, and assuming a $kT\sim 6$ keV, they derived
from a {\em ROSAT} count rate of 0.02211 counts s$^{-1}$ an
X-ray luminosity of $L_{X({\rm 0.5-2.0 keV})}=2.87 \times 10^{44}$  erg~s$^{-1}$
(H$_0$=50~km~s$^{-1}$~Mpc$^{-1}$ and $q_0 = 0.5$).

\underline{1WGA~J0432.4$+$1723}\\	
\label{1WGAJ043241723}		
This source was found by Carkner et al. (1996) in the outer PSPC region
 (offset $\sim 37^{\prime}$) of a 1993 observation (rp900353n00) of 
Lynds 1551, a well-studied cloud in the Taurus-Auriga star forming region.
They found a 7$\sigma$ source with count rate of $1.9 \times 10^{-2}$ counts
s$^{-1}$.
Even though they do not make a clear identification, the source is included 
among the sample of possible T~Tauri stars. 
The authors do not find any evidence near 1WGA~J0432.4$+$1723 
for $H{\alpha}$ emission or 
Li 6707 \AA \/ absorption, typical of T-Tauri stars. Moreover 
the proper motion 
 they measured is inconsistent with the source  being part of the cloud.
They conclude 
that 1WGA~J0432.4$+$1723 is likely to be unrelated to the cloud itself.
1WGA~J0432.4$+$1723 was also detected, but not classified, in the analysis of 
the RASS by Wichmann et al. (1996).\\
In a  1991 observation, the source
falls in the inner PSPC detector region. The $\sim 2.5$ times longer
exposure  time  and the sharper PSF in the inner circle give more
precise results than in Carkner et al. (1996). We find a (0.5-2.0 keV) 
count rate of $(1.42 \pm 0.1) \times 10^{-2}$ counts s$^{-1}$ 
and a position (04:32:29.5, $+$17:23:45.4) consistent with the Galpipe.
The fit of the spectrum with a black body model suggests a temperature of $\sim
0.55 \pm 0.01$~keV, somewhat lower than the $\sim 1$ keV expected for a 
T Tauri star  but not conclusive due to the patchy nature of the ISM 
in this area.
A NVSS radio source with flux  equal to $3.2 \pm 0.6$ mJy is
present $\sim 17^{\prime \prime}$ from the PSPC position.
Radio continuum emission is often observed from  T-Tauri stars at a
level of 10$^{15}$ - 10$^{18}$ erg s$^{-1}$ Hz$^{-1}$ and, 
assuming a distance of 140 pc for L1551, similar to the other Taurus-Auriga
 star forming clouds (e.g. Carkner et al. 1996), 
the radio flux corresponds
to a luminosity $\sim 10^{17}$~erg s$^{-1}$ Hz$^{-1}$.
However the proper motion 
as reported in Carkner et al. (1996) exclude the possibility 
of a new T-Tauri star.
 1WGA~J0432.4$+$1723 was observed by ROSAT in 1991 and in 1993.
The source is persistent 
and does not show evidence for variability on a 6-hour timescale 
(the longest  timescale sampled, corresponding to the 1991 
observation)\footnote{In 1993 the source is detected in the outer rim 
and spread over a large
 area and a variability study is not possible due to the low signal to noise. 
The count rate, however, is consistent with that of the 1991 observation}.

1WGA~J0432.4$+$1723 has an E=20.0 Palomar counterpart in its error circle;
for  O$>21.5$, the counterpart is at least moderately red ($O-E>1.5$).
We propose a high redshift cluster classification for 1WGA~J0432.4$+$1723
on the basis of the
results of follow-up optical and IR imaging observations we performed 
(Cagnoni et al., in preparation).
These observations show the presence of  an IR-bright 
extremely red galaxy in the error circle, and  an excess of similarly 
red sources nearby (see also Cagnoni et al. 2001b, 1WGA~J0432.4$+$1723
is the  candidate high z cluster mentioned therein). 


\underline{1WGA~J1103.5$+$2459}\\		
\label{1WGAJ110352459}	
1WGA~J1103.5$+$2459 is listed in Romer et al. (2000) as an extended source 
detected in the SHARC survey with a (wavelet) count rate of 0.00411 
counts s$^{-1}$, but is not  identified and is not included in the 
SHARC cluster sample. 
This source has a red Palomar counterpart (E=18.2, which implies O-E$>3.3$) 
in its error circle and could thus be 
another high redshift cluster of galaxies.


\subsubsection{Discussion on clusters of galaxies}
\label{blanks_clusters}

The unabsorbed luminosities of the two high redshift  clusters of galaxies
we found are high:
$L_{X(0.5-2.0)}=1.28 \times 10^{44}$  erg~s$^{-1}$
for  1WGA~J0221.1$+$1958, Romer et al. (2000);  
and $L_{X(0.5-2.0)}=(4.4 \pm 0.5) \times 10^{44}$ erg~s$^{-1}$
for 1WGA~J1226.9$+$3332,  Cagnoni et al. (2001a).
Since high luminosity, high redshift clusters should be rare, 
the relative ease with which we discovered them
is potentially of great  significance.  
The search for extremely X-ray loud sources  can sample 
a large area of the sky and 
while the high cut on the sources flux includes the brightest clusters only,
the optical cut removes the low z ones.
As a result only the few interesting high redshift 
and high luminosity clusters of galaxies candidates are selected 
among the $\sim 62,000$ sources in the WGACAT95.
With  two high 
z and bright clusters of galaxies already found out of the 16 sources selected in the WGACAT
make this method
the most efficient way to select these rare sources. 

\subsection{Ultraluminous X-ray sources in nearby galaxies}
\label{blanks_ulx}

\underline{1WGA~J1243.6$+$3204}\\		
\label{1WGAJ124363204}	
This source was detected  by Vogler, Pietsch \&  Kahabka  (1996) in the same
observation as the brightest
{\it ROSAT} source within the optical extent of the edge-on spiral
galaxy NGC~4656 (z=0.00215).
It is located to the south-west, at $\sim 7.6^{\prime}$ 
(corresponding to 0.3~kpc at NGC~4656 redshift) from the
nucleus along the major axis.
In this  region there is a deficiency in diffuse, galaxy related 
X-ray emission, probably due to the cold gas seen in H{\sc I} 
which bridges the space 
between NGC~4656 and its tidally interacting companions 
(Vogler, Pietsch \&  Kahabka  1996).
These authors  estimated 
 a 0.1-2.4 keV count rate of 
 $(9.3 \pm 0.8) \times 10^{-3}$ counts s$^{-1}$, corresponding to
 $f_X \sim 1.3 \times 10^{-13}$ erg cm$^{-2}$
s$^{-1}$ (0.1-2.4 keV).
The N$_H$ value from their fit
is  in excess of  $5 \times 10^{21}$ cm$^{-2}$,
 both for a thin thermal plasma and for a power law model.
The absorbing column is larger than the H{\sc i} density of 
$\sim 8 \times 10^{20}$cm$^{-2}$ within
NGC~4656 in this direction. There is a 0.3\% probability for this
blank  to be a background X-ray source (Vogler, Pietsch \&  Kahabka  1996).
They also investigate the possibility that 1WGA~J1243.6$+$3204 is
 a heavily absorbed source within NGC~4656, e.g. a X-ray binary.
In this case, $L_X \sim 8 \times 10^{38}$ erg s$^{-1}$ and the source
would be radiating well above the Eddington limit for a one solar mass
system and need a 10 $M_{\odot}$ black hole to power it.  
We find a position  and a count rate  consistent with Vogler, Pietsch \&  Kahabka
 (1996).
Fitting the spectrum with an absorbed power law in the range 0.07 --
2.4 keV, we find an absorbing column of $1.71^{+1.50}_{-0.78} \times
10^{21}$ cm$^2$, clearly in excess of the Galactic absorption 
($\sim 1.23 \times 10^{20}$ cm$^{-2}$) and of the  column density within
NGC~4656  in this direction 
($8 \times 10^{20}$cm$^{-2}$).
 1WGA~J1243.6$+$3204 was also serendipitously observed in the {\it ROSAT} HRI 
in a $\sim 27$ ks long observation performed in 1994, two years after
the PSPC one. From the HRI image the source appears pointlike with a
more accurate position (12:43:41.1, +32:04:56.6) and a count rate of $(3.4
\pm 0.6 ) \times 10^{-3}$ counts s$^{-1}$.
The HRI and PSPC count 
rates agree within $1\sigma$.\\
The same HRI image was also analyzed by Roberts \& Warwick (2000)
and 1WGA~J1243.6$+$3204 is  detected as a serendipitous source with a 
count rate of $(3.0 \pm 0.4) \times 10^{-3}$ counts s$^{-1}$.


\underline{1WGA~J1216.9$+$3743}\\
1WGA~J1216.9$+$3743, as 1WGA~J1243.6$+$3204, lies along the major axis  of a nearby galaxy (NGC~4244, at 3.6 Mpc), at $\sim 6^{\prime}$ ($\sim 6$~kpc) off the galaxy center.  
1WGA~J1216.9$+$3743 X-ray emission, as 1WGA~J1243.6$+$3204 one,  
is strongly absorbed  and extremely soft
($\Gamma \sim 5$, see Table~\ref{blanks_tab_fits}). 
1WGA~J1216.9$+$3743 was not detected by the HRI, as expected from its PSPC count rate.
For a {\em Chandra} observation of this source, 
which confirms the extremely soft ULX nature, see Cagnoni et al. (2002).

\subsubsection{ULX and X-ray binaries discussion}

We suggest that these sources could be X-ray binaries in   nearby galaxies, as already proposed by Vogler, Pietsch \&  Kahabka  (1996) for 1WGA~J1243.6$+$3204
(see Figure~\ref{blanks_binaries}). 
If this is the case, 
they would be radiating above the Eddington limit for 1$M_{\odot}$ star
(unabsorbed $L_{(\rm 0.5-2.0 keV)} \sim 10^{41}$ and  $\sim 10^{39}$ erg s$^{-1}$ rispectively).
These objects in fact appear similar to MS~0317.7$-$6647, the brightest unidentified object of the EMSS (Stocke et al. 1995),
and to the ultraluminous X-ray sources which are now being found by {\em Chandra}
(e.g. Fabbiano, Zezas \& Murray 2001; Blanton, Sarazin \& Irwin, 2001; Sarazin, Irwin \& Bregman, 2000; 2001; Cagnoni et al. 2002).



\subsection{Unknown nature}
\label{blanks_individual_unknown}

\underline{1WGA~J1220.6$+$3347}\\		
\label{1WGAJ122063347}	
1WGA~J1220.6$+$3347 was detected by the {\it Einstein}  
IPC as an ultra-soft source
(Thompson, Shelton \&  Arning 1998)  with
$CR =0.012 \pm  0.002$ counts s$^{-1}$ more than 10 years before the
PSPC observation. 
{\em Einstein} CR is consistent within the errors with the predisctions based
on the PSPC CR.
The {\it Einstein} and Galpipe PSPC positions are 22$^{\prime \prime}$ apart
and so are consistent.\\

\underline{1WGA~J1233.3$+$6910}\\		
\label{1WGAJ123336910}	
This  source is listed in Radecke (1997)
with a  position of 12:33:23.69 +69:10:05.5 and a count rate of $0.0225$ 
counts s$^{-1}$.
For the same observation we found a position $20.4^{\prime \prime}$ away 
(Table~\ref{blanks_tab_16src}) and a count rate of
$0.0214 \pm 0.0013$ counts~s$^{-1}$.
5 years after the PSPC observation, the source was detected by the HRI; 
however  the source falls on the detector edge, 
making any flux and positional estimate unreliable.\\

\underline{1WGA~J1420.0$+$0625}\\
1WGA~J1420.0$+$0625 was observed twice by {\em ROSAT} and is consistent
with no variability  within each  observation and  on a 6-month timescale.

\subsubsection{Discussion on the unknown sources}

The three sources presented above, besides being all persistent, 
have similar unexceptional 
X-ray spectral shapes ($\Gamma \sim 2.2 -2.4$), $N_H = N_{H{\rm (Gal.)}}$ (Table~\ref{blanks_tab_fits}), lack a red Palomar counterpart 
and have $f_X/f_V \geq 15$ (Table~\ref{blanks_fx}).
Could we be seeing the reflected component  of an instrinsecally bright
 obscured AGN?

\section{Comparison with previous surveys}

A number of different projects to identify X-ray-discovered sources have
been carried out over the last decade. Although the major goal of these
studies was not to search for blanks, they were nevertheless 
discovered as a few percent `remnant' of unidentified sources (e.g. Bade 
et al. 1995; Schwope et al. 2000, and references therein).
For instance, Schwope et al. (2000 and references therein) presented a 
large sample of optically (un)identified {\em ROSAT} sources, 
in which there were three cases of blanks
\footnote{These 3 blanks are not included in our sample because they
were detected only with the HRI, or with the PSPC data after the 
release ofWGACAT95.}
.
 Blanks usually consitute $\sim 5-8$\%  of all wide-angle surveys, 
regardless of survey depth. Because their optical faintness prevents 
easy investigation, blanks have not been investigated  further in these 
general purpose wide-angle surveys. 

Blanks are also found in deep fields (e.g. CDFN, Alexander et al. 2001; 
UDS, Lehmann et al. 2000), and these are usually investigated in 
greater
detail than in wide-angle surveys, despite being even more difficult to
follow-up, because of the great interest in deep survey identifications in 
general.

Wide-angle surveys might be expected to trace different populations than
narrow-angle deep fields. Yet a comparison of our identifications with those
of the CDFN, UDS and UK-RDS shows that they overlap quite strongly.
The counterparts of blanks found in these deep fields are usually 
red. The favored classifications are high redshift obscured quasars, and
high redshift  ($z\sim 1$) clusters of galaxies (e.g. Newsam et al. 1997; 
Alexander et al. 2001; Lehmann et al. 2001). This makes the bright blanks from 
our, and other, wide-field surveys of immediate interest for understanding
the faint source population. Being 100-1000 times brighter in X-rays, they
are far easier to study in detail. As a result we anticipate that they will 
become intensely studied.

\section{Summary and conclusions}
\label{blanks_summary}

We have identified a population of \bfs \/  among the {\em ROSAT} 
bright X-ray sources with faint optical 
counterparts, i.e. $O>21.5$ on the Palomar Sky survey.
Their extreme X-ray over optical flux ratio (\fxv $> 10$) 
is not compatible with 
the main classes of X-ray emitters, except for BL Lacs 
for the less extreme cases.
The identification process brought to the discovery of two high z, high 
luminosity clusters  of galaxies (\S ~\ref{blanks_clusters}), 
one BL Lacertae object (\S~\ref{1WGAJ134012743}), and two
type~1 AGNs (\S ~\ref{blanks_agns}).
Four  blanks are  AGN-like sources 
(\S ~\ref{blanks_agns}) which seem to form a well defined group:
they present  type~1 X-ray spectra and red Palomar counterparts. 
These sources are similar to the galaxies with bright X-ray emission 
($L_X \sim 10^{42} - 10^{43}$ erg s$^{-1}$)
but weak or absent AGN features in the optical band found since
 {\em Einstein} observations (Elvis et al. 1981; Tananbaum et al. 1997) and 
which are  now being found in large numbers 
(e.g. Della Ceca et al. 2001; Hornschemeier et al. 2001, Fiore et al. 2001).
We considered possible explanations for
the discrepancy between  X-ray and optical data: a
suppressed BBB emission; 
an extreme 
dust to gas ratio;
a dust grain size distribution different from the Galactic one;
a dusty warm absorber and  an high redshift ($z \geq 3.5$) QSO nature.
Two sources (\S ~\ref{blanks_ulx}) are candidate ultraluminous X-ray binaries
within nearby galaxies (see also Cagnoni et al. 2002).
Three sources (\S ~\ref{blanks_individual_unknown}) have a still unknown nature; for each 
of them we listed and justified the possibilities excluded.\\

To make progress in understanding the nature of blanks
we need to identify them with optical or near infrared counterparts.
Since both in the case of obscured AGNs and of high redshift clusters 
of galaxies red counterparts are expected, we
 obtained optical (R band) and infrared (K band) imaging for the 
16 fields in our sample at a 4m-class telescope and we will present the
results in a future paper.\\
The obvious next step is to construct a larger and statistically complete sample 
of \bfs . For this purpose, the {\em XMM-Newton} and {\em Chandra}
 catalogs soon available 
will form a good basis. The smaller positional uncertainty  
will cut down on false optical identifications and so will considerably 
{\em increase} the percentage of blanks found ($\sim 14$\% at $f_X \geq 10^{-13}$ \cgs ,  Maccacaro \& Della Ceca private communication). 
The smaller PSF of these new detectors
will also allow the direct separation of extended
sources, enabling high redshift clusters of galaxies to be found even more 
readily.

\acknowledgments

\begin{figure}
\plotone{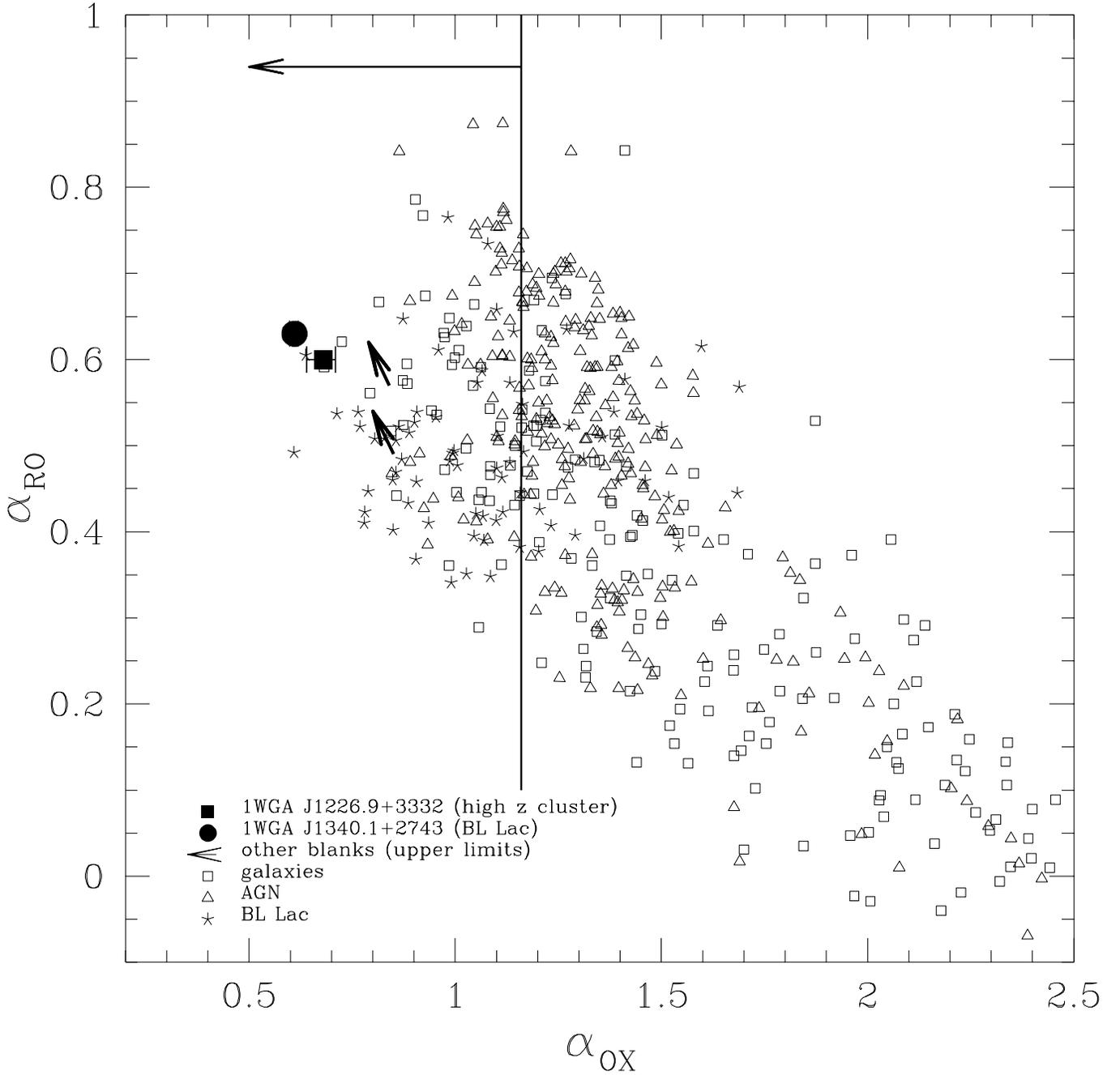}
\caption{\footnotesize{Broad band spectral indices for the 4 \bfs \/ with a 
radio counterpart compared to Caccianiga et al. (1999) sources
(BL Lacs are represented by stars, 
emission line AGNs by open triangles and galaxies by open squares).
The two filled symbols and the two nearby  arrows represent 
the 4 blanks with a radio counterpart;
in particular the square corresponds to the high z cluster 1WGA~J1226.9$+$3332,
the circle to the BL Lac object 1WGA~J1340.1$+$2743 and the two arrows represent 
the limits for 1WGA~J0432.4$+$1723 and 1WGA~J0221.1$+$1958. 
The vertical line indicates the highest $\alpha_{ox}$ in our sample.}\label{blanks_caccia}}
\end{figure}

\begin{figure}
\plotone{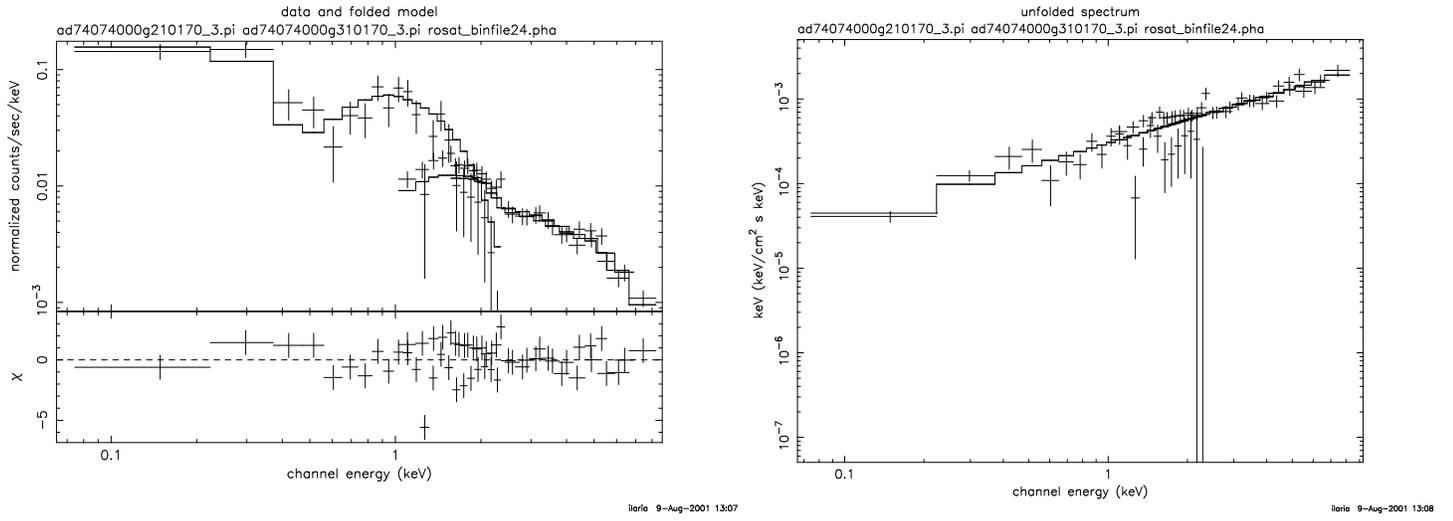}
\caption{\footnotesize{Left: {\em ROSAT} PSPC and {\em ASCA} GIS2 and GIS 3 
spectra of  1WGA~J1220.3$+$0641 and residuals to the fit with a powerlaw 
with absorption fixed to the Galactic value. Right: {\em ROSAT -- ASCA} 
spectrum in the $\nu - F\nu$ plane.}\label{blanks_asca_sp}}
\end{figure}

\begin{figure}
\plotone{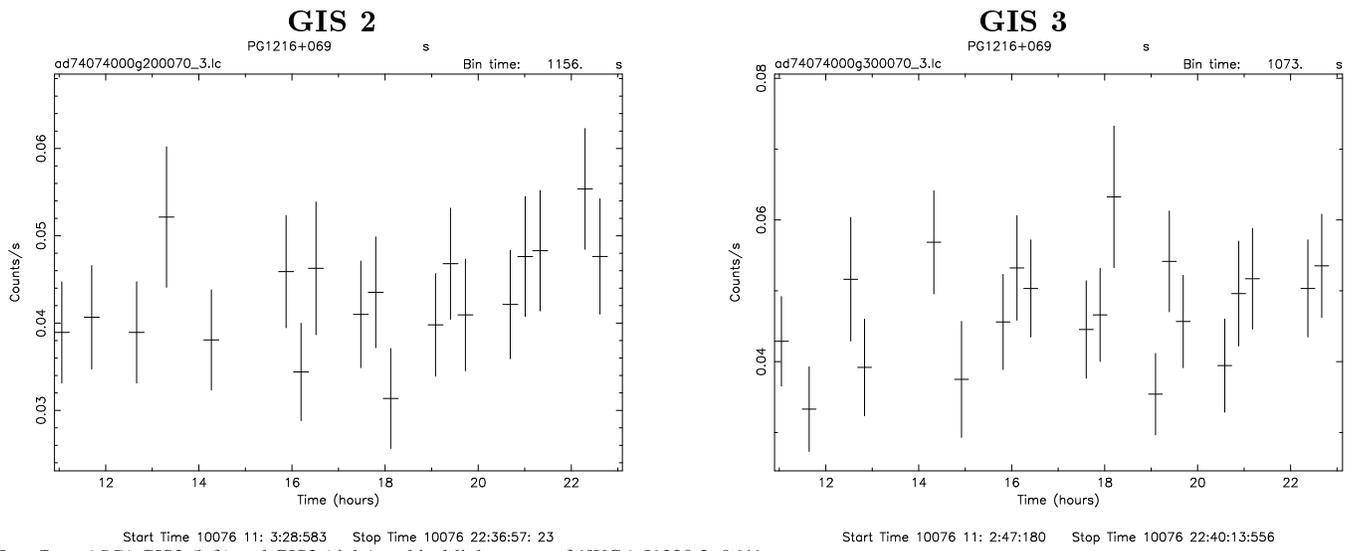}
\vspace{0.2cm}
\caption{\footnotesize{{\em ASCA} GIS2 (left) and GIS3 (right) archival lightcurves 
of 1WGA~J1220.3$+$0641.}\label{blanks_asca_lc}}
\vspace{2cm}
\footnotesize{FIG. 8 ---POSS~{\sc ii} blue images of (left)  NGC~4656 with the circle 
indicating 1WGA~J1243.6$+$3204
position and (right) of NGC~4244 with the circle indicating the 
1WGA~J1216.9$+$3743  position (North is up and east is on the left; 
image sizes are $15^{\prime}  \times 15^{\prime}$ for NGC~4656 and  
$20^{\prime}  \times 20^{\prime}$ for NGC~4244).}
\end{figure}

This work made use of: the NASA/IPAC Extragalactic Database (NED) which is operated by the 
Jet Propulsion 
Laboratory, California Institute of Technology, under contract with the National 
Aeronautics and Space Administration; the HEASARC archive, a service of the Laboratory for 
High Energy 
Astrophysics (LHEA) at NASA/ GSFC and the High Energy Astrophysics Division of the 
Smithsonian  Astrophysical Observatory (SAO); 
the SIMBAD database, operated at CDS, Strasbourg, France.\\
I.C. thanks Fabrizio Fiore for making his program to compute the 
effective spectral indices available, Alessandro Caccianiga
for the electronic format of his data used for Figure~\ref{blanks_caccia}
and Roberto Della Ceca,  Paola Severgnini and Aldo Treves for
useful scientific discussions.
We thank the anonymous referee for the carefull and deep work and for 
 helpful comments and suggestions,
which improved the quality of this paper.
This work was supported by NASA ADP grant NAG5-9206 
and by the Italian MIUR (IC and AC).
I.C. acknowledges a CNAA fellowship.

\newpage

\begin{deluxetable}{l c c c c c c c }
\footnotesize
\tablewidth{7.2in} 
\tablecaption{The final sample of 16 sources.}
\tablehead{
\colhead{Source name}
& \colhead{Best coords\tablenotemark{a}}
& \colhead{PSPC}
& \colhead{HRI}
& \colhead{POSS}
& \colhead{POSS}
& \colhead{POSS}
& \colhead{$F_X/F_V$}\\
\colhead{(1WGA~J)} 
& \colhead{Ra, Dec (J2000)}
& \colhead{count rate\tablenotemark{b}}
& \colhead{count rate\tablenotemark{c}}
& \colhead{E}
& \colhead{O}
& \colhead{O--E\tablenotemark{d}}
&
}
\startdata
0221.1$+$1958	&02 21 09.0,  19 58 15.3 	&$1.29 \pm 0.09$ 	&--	&- 		&- 	 	&-	&$>$21.3\\
0432.4$+$1723	&04 32 29.6,  17 23 47.8 	&$1.42 \pm 0.10$ 	&--	&20.0		&- 	 	&$>1.5$	&$>$38.5\\
0951.4$+$3916	&09 51 28.7,  39 16 36.8 	&$0.90 \pm 0.10$ 	&--	&19.1 	&- 		&$>2.4$	&$>$13.0\\
1103.5$+$2459	&11 03 35.4,  24 59 09.6 	&$0.33 \pm 0.04$ 	&--	&18.2 	&-  		&$>3.3$	&$>$3.5\\
1216.9$+$3743	&12 16 57.1,  37 43 35.4 	&$1.26 \pm 0.15$ 	&$<0.23$\tablenotemark{f}	&-  	&- 	 	&-	&$>$13.5\\
1220.3$+$0641	&12 20 18.3,  06 41 25.7\tablenotemark{e}		&$4.42 \pm 0.39$ 	&$2.94 \pm 0.53$	&18.58 	&21.81 		&3.23	&102.1\\
1220.6$+$3347	&12 20 38.4,  33 47 26.7 	&$0.65 \pm 0.07$ 	&--	&20.9 	&21.7 		&0.8	&14.6\\
1226.9$+$3332	&12 26 57.6,  33 33 00.9\tablenotemark{e}		&$2.00 \pm 0.12$ 	&$0.87 \pm 0.15$	&- 	&- 		&-	&$>$41.4\\
1233.3$+$6910	&12 33 25.5,  69 10 14.6 	&$2.14 \pm 0.13$ 	&on the border	&- 	&- 		&-	&$>$38.4\\
1243.6$+$3204	&12 43 41.1,  32 04 56.6\tablenotemark{e}		&$0.67 \pm 0.08$ 	&$0.34 \pm 0.06$	&- 	&- 		&-	&$>$9.7\\
1340.1$+$2743	&13 40 10.3,  27 43 38.9 	&$4.74 \pm 0.23$ 	&--	&- 	&- 		&-	&$>$83.4\\
1412.3$+$4355	&14 12 21.4,  43 55 01.0\tablenotemark{e}	&$0.53 \pm 0.06$ 	&$0.19 \pm 0.09$	&18.73 	&- 		&$>2.77$	&$>$9.0\\
1415.2$+$4403	&14 15 15.0,  44 03 20.2 	&$0.30 \pm 0.06$ 	&--	&- 	&- 		&-		&$>$6.0\\
1416.2$+$1136	&14 16 13.2,  11 36 17.9\tablenotemark{e}		&$0.31 \pm 0.08$ 	&$0.21 \pm 0.09$	&19.5    &22.22 	 &2.72	&10.7\\
1420.0$+$0625	&14 20 05.6,  06 25 25.0 	&$0.85 \pm 0.09$ 	&--	&- 		&- 		&-	&$>$15.7\\
1535.0$+$2336	&15 35 06.1,  23 37 03.2\tablenotemark{e}		&$0.37 \pm 0.07$ 	&$0.14 \pm 0.03$	&19.6 	&21.8 	 &2.2	&10.2\\
\enddata
\label{blanks_tab_16src}
\tablenotetext{a}{Best X-ray coordinates: Galpipe ($\pm 13^{\prime \prime}$ 1$\sigma$) or HRI ($\pm 1\sigma$) when available.}
\tablenotetext{b}{between 0.5 and 2.0 keV in units of $10^{-2}$ counts s$^{-1}$.}
\tablenotetext{c}{$10^{-2}$ counts s$^{-1}$.}
\tablenotetext{d}{Assuming O magnitude equal to V magnitude and O$>$21.5 when no source is detected in the Palomar O band}
\tablenotetext{e}{HRI coordinates}
\tablenotetext{f}{$3 \sigma$ upper limit in a $r=24^{\prime \prime}$ circle}
\end{deluxetable}

\newpage

\begin{deluxetable}{c c c c c c c c}
\footnotesize
\tablewidth{7.0in} 
\tablecaption{All available {\em ROSAT} observations for the 16 blank field sources.}
\tablehead{
\colhead{}
&\multicolumn{4}{c}{PSPC observations}
&\multicolumn{3}{c}{HRI observations}\\
\colhead{(1WGA~J)} 
& \colhead{ROR} 
& \colhead {Date}
& \colhead {Exposure}
& \colhead {Counts\tablenotemark{a}}
& \colhead {ROR}
& \colhead {Date}
& \colhead{Exposure}
}
\startdata
0221.1$+$1958	&900147n00\tablenotemark{c}	&07-30-1991	&24947	&365	&--	&--	&--\\
\hline
0432.4$+$1723	&200443n00\tablenotemark{c}	&03-07-1991	&20074	&313	&--	&--	&--\\
		&201313n00	&09-10-1992	&4027	&--	&--	&--	&--\\
		&900353n00	&02-21-1993	&7718	&--	&--	&--	&--\\
\hline
0951.4$+$3916	&701367n00	&10-26-1993	&14281	&116	&--	&--	&--\\
\hline
1103.5$+$2459	&300291n00\tablenotemark{c}	&05-26-1993	&44010	&357	&--	&--	&--\\
\hline
1216.9$+$3743	&600179n00	&11-21-1991	&9304	&135	&702724n00	&06-20-1996	&8752\\
\hline
1220.3$+$0641	&700021n00\tablenotemark{c}	&12-20-1991	&3430	&337	&702172n00	&12-10-1995	&2348\\
		&\ \ \ \ \ --			&--	&--		&--	&702173n00	&12-11-1996	&2662\\
		&\ \ \ \ \ --			&--	&--		&--	&702174n00	&12-12-1995	&3183\\
		&\ \ \ \ \ --			&--	&--		&--	&702188n00	&06-07-1996	&3951\\
		&\ \ \ \ \ --			&--	&--		&--	&702189n00	&07-06-1996	&3835\\
		&\ \ \ \ \ --			&--	&--		&--	&702210n00	&12-09-1995	&2476\\
		&\ \ \ \ \ --			&--	&--		&--	&702211n00	&12-17-1995	&3839\\
		&\ \ \ \ \ --			&--	&--		&--	&702212n00	&12-23-1995	&890\\
\hline
1220.6$+$3347	&700864n00	&06-20-1992	&3009	&427\tablenotemark{b}	&--	&--	&--\\
		&700864a01	&05-23-1993	&19284	&427\tablenotemark{b}	&--	&--	&--\\
\hline
1226.9$+$3332	&600173n00	&06-02-1992	&8002	&538\tablenotemark{b}	&702725n00	&06-23-1996	&11353\\
		&600277n00	&06-17-1992	&9036	&538\tablenotemark{b}	&--	&--	&--\\
\hline
1233.3$+$6910	&300034n00	&04-04-1991	&15040	&671	&300492n00	&10-08-1996	&41834\\
\hline
1243.6$+$3204	&600416n00\tablenotemark{c}	&06-28-1992	&18052	&103	&600605n00	&05-29-1994	&27669\\
\hline
1340.1$+$2743	&300333n00	&07-06-1993	&13083			&2220\tablenotemark{b}	&--	&--	&--\\
		&300333a01	&07-02-1994	&14599			&2220\tablenotemark{b}		&--	&--	&--\\
		&701063n00	&07-15-1992	&10032			&2220\tablenotemark{b}		&--	&--	&--\\
		&701065n00	&07-11-1992	&8971			&2220\tablenotemark{b}		&--	&--	&--\\
		&701067n00	&07-14-1992	&8209			&2220\tablenotemark{b}		&--	&--	&--\\
		&701069n00\tablenotemark{c}	&07-13-1992	&9912	&822	&--	&--	&--\\
		&701459n00	&07-06-1993	&7157			&2220\tablenotemark{b}		&--	&--	&--\\
\hline
1412.3$+$4355	&700248n00	&06-26-1991	&24843	&306	&--	&--	&--\\
\hline
1415.2$+$4403	&700248n00	&06-26-1991	&24843	&322	&--	&--	&--\\
\hline
1416.2$+$1136	&700122n00	&07-20-1991	&27863	&265	&701858n00	&07-16-1995	&11345\\
\hline
1420.0$+$0625	&700865n00	&07-25-1992	&10710	&268\tablenotemark{b}	&--	&--	&--\\
		&700865a01	&01-12-1993	&8876	&268\tablenotemark{b}     &--	&--	&--\\
\hline
1535.0$+$2336	&701411n00\tablenotemark{c}	&07-24-1993	&22955	&130	&701330n00	&07-29-1994	&35661\\
\enddata
\label{blanks_pspc_hri}
\tablenotetext{a}{Source net counts in each observation between 0.07 and 2.4 keV}
\tablenotetext{b}{Calculated in the combined image of all the available PSPC observations}
\tablenotetext{c}{WGA observation}
\end{deluxetable}

\newpage

\begin{landscape}
\begin{deluxetable}{ c| c c c c c | c c c c | c c c c| }
\footnotesize
\tablewidth{10in}
\tablecaption{Spectral fits with: an absorbed power law, a black body and an absorbed Raymond-Smith model}
\tablehead{
\multicolumn{1}{c}{Source name} 
&\multicolumn{5}{c}{Absorbed power law} 
&\multicolumn{3}{c}{Absorbed Black body}
&\multicolumn{4}{c}{Absorbed Raymond-Smith}\\
\colhead{1WGA~J}
&\colhead{$\Gamma$\tablenotemark{a}}
&\colhead{k\tablenotemark{b}}
&\colhead{$N_H$\tablenotemark{c}}
&\colhead{$N_{H(gal)}$\tablenotemark{d}}
&\colhead{$\chi_r^2$\tablenotemark{e}}
&\colhead{kT\tablenotemark{f}}
&\colhead{k\tablenotemark{b}}
&\colhead{$N_H$\tablenotemark{c}}
&\colhead{$\chi_r^2$\tablenotemark{e}}
&\colhead{kT\tablenotemark{g}}
&\colhead{z}
&\colhead{k\tablenotemark{b}}
&\colhead{$\chi_r^2$\tablenotemark{e}}
}
\startdata
0221.1$+$1958 &$2.33^{+0.65}_{-0.47}$	&$0.96^{+0.34}_{-0.16}$	&$12.29^{+9.03}_{-5.23}$ &9.59 &0.96	
&$303^{+38}_{-15}$	&$2.23^{+0.25}_{-0.19}$ &$2.62^{+4.19}_{-1.91}$	 &1.16
&$2.23^{+1.21}_{-0.57}$	&$0.49^{h}_{-0.38}$	&$7.91^{+16.76}_{-7.91}$ &0.96\\
&$2.15 \pm 0.28$		&$9.21^{+0.37}_{-1.00}$ &fixed 	
		&--   &0.93 & & & & & & & &\\
0432.4$+$1723 &$1.03^{+0.68}_{-0.46}$	&$0.92^{+0.44}_{-0.17}$	&$9.42^{+12.40}_{-5.34}$	&13.1 &0.77
&$496^{+108}_{-82}$	&$3.88^{+1.14}_{-0.60}$ &$2.72^{+6.83}_{-2.35}$ &0.84
&$0.32^{+0.09}_{-0.05}$	&$1.98^{h}_{-0.77}$ &$942^{+26.29}_{-7.26}$ &1.29\\ 
&$1.20^{+0.23}_{-0.26}$	&$10.56^{+6.29}_{-1.33}$ &fixed 
			&--   &0.95 & & & & & & & &\\
0951.4$+$3916 &$2.62^{+2.14}_{-0.81}$	&$0.82^{+1.72}_{-1.64}$ &$20.19^{+3.99}_{-1.09}$ &1.55 &0.93
&$295^{+65}_{-60}$	&$1.71^{+0.93}_{-0.29}$	&$7.40^{+16.26}_{-5.85}$	&0.94
&$64.00^{h}_{-52.68}$	&$0.01^{h}_{h}$	&$2.93^{+9.42}_{-2.93}$	&1.40\\
	      &$0.92^{+0.25}_{-0.28}$	&$4.21^{+0.39}_{-0.67}$ &fixed 	
		&--   &1.29 & & & & & & & &\\
1103.5$+$2459 &$1.67 \pm 0.31$		&$0.16^\pm 0.02$	&$0.35^{+0.82}_{-0.29}$ &1.38 &0.92
&$222^{+64}_{-49}$	&$0.55^{+0.09}_{-0.07}$	&$0.0^{+0.15}_{h}$	&1.69
&$2.84^{+1.49}_{-0.93}$	&$2.00^{h}_{-1.22}$	&$6.72^{+0.77}_{-6.72}$ &1.17\\
	      &$2.10^{+0.17}_{-0.18}$	&$1.76^{+0.20}_{-0.25}$ &fixed 	
		&--   &0.94 & & & & & & & &\\
1216.9$+$3743 &$4.90^{+5.10}_{-1.74}$	&$2.17^{+4.35}_{-19.43}$	&$44.78^{+8.72}_{-2.43}$ &1.69 &0.81
&$194^{+82}_{-77}$	&$3.41^{+54.51}_{-3.41}$	&$14.94^{+48.54}_{-12.62}$	&0.82
&$0.77^{+0.13}_{-0.10}$	&$0.00^{+0.04}_{h}$	&$1.18^{+0.42}_{-0.33}$	&0.96\\
&$1.34^{+0.25}_{-0.31}$	&$5.76^{+0.40}_{-1.27}$ &fixed 	
		&--   &1.16 & & & & & & & &\\
1220.3$+$0641 &$1.95^{+0.28}_{-0.25}$	&$2.22^{+0.27}_{-0.17}$  &$1.28^{+0.66}_{-0.54}$	&1.56 &1.34	
&$174$ &$7.99$	&$3.93 \times 10^{-8}$ &2.67
&$2.52^{+0.62}_{-0.42}$	&$2.00^{h}_{-0.68}$	&$97.3^{+3.3}_{-97.3}$	&1.67\\
	      &$2.07 \pm 0.10$		&$23.39^{+1.67}_{-2.66}$ &fixed 
			&--   &1.29 & & & & & & & &\\
1220.6$+$3347 &$2.31 \pm 0.33$		&$0.37^{+0.03}_{-0.06}$	&$1.40^{+0.29}_{-0.68}$	&1.27 &0.63
&$147^{+8}_{-7}$	&$1.52 \pm 0.12$	&$0.0^{+0.10}_{h}$	&1.63
&$1.77^{+0.35}_{-0.47}$	&$2.00^{h}_{-0.69}$	&$20.8^{+2.4}_{-20.8}$	&0.88\\
	      &$2.26 \pm 0.11$		&$3.70^{+0.29}_{-0.59}$ &fixed 	
		&--   &0.60 & & & & & & & &\\
1226.9$+$3332 &$1.51^{+0.24}_{-0.23}$	&$1.06^{+0.12}_{-0.05}$  &$1.66^{+0.84}_{-0.64}$	&1.38 &1.08	
&$300^{+30}_{-26}$	&$3.37^{+0.24}_{-0.22}$	&$0.0^{+0.09}_{h}$	&1.93
&$7.42^{+9.18}_{-2.81}$	&$0.56^{h}_{h}$	&$8.89^{+16.97}_{-8.89}$	&1.07\\
&$1.42 \pm 0.11$		&$11.05^{+0.40}_{-0.98}$ &fixed 
			&--   &1.04 & & & & & & & &\\

1233.3$+$6910 &$2.43^{+0.20}_{-0.18}$	&$1.17^{+0.07}_{-0.09}$	&$2.60^{+0.57}_{-0.53}$	&1.69 &1.39	
&187	&4.04	&$1.12 \times 10^{-10}$	&2.5
&$2.23^{+0.53}_{-0.50}$	&$1.75^{h}_{-0.45}$ &$43.4^{+8.7}_{-43.4}$ &1.47\\
      &$2.12 \pm 0.08$		&$11.39^{+0.50}_{-1.03}$	&fixed 	
		&--   &1.45 & & & & & & & &\\

1243.6$+$3204 &$5.97^{+6.62}_{-2.59}$	&$18.3^{+366.7}_{-36.7}$	&$171.3^{+149.5}_{-78.3}$ &1.23 &0.49
&$770^{+40}_{-58}$	&$11.33^{+210.38}_{-11.33}$	&$110.5^{+109.5}_{-49.0}$	&0.48	
&$64.0^{h}_{-30.76}$	&$0.0^{+0.89}_{h}$ &$2.26^{+2.65}_{-2.26}$ &1.76\\
&$0.07^{+0.22}_{-0.53}$	&$3.14^{+0.35}_{-0.61}$ &fixed 	
		&--   &0.84 & & & & & & & &\\

1340.1$+$2743 &$2.35 \pm 0.18$		&$1.18^{+0.09}_{-0.04}$  &$2.99^{+0.58}_{-0.54}$	&1.09 &0.53	
&$213$	&$8.35$	&$4.04 \times 10^{-6}$	&2.15
&$2.69$	&$0.46$	&$20.0$	&2.04\\
	      &$1.72 \pm 0.06$		&$10.83^{+0.42}_{-0.55}$ &fixed 
			&--   &1.25 & & & & & & & &\\

1412.3$+$4355 &$1.76^{+0.30}_{-0.30}$	&$0.25 \pm 0.03$	&$0.06^{+0.09}_{-0.03}$ &1.16 &0.48
&$168^{+21}_{-16}$	&$0.93 \pm 0.11$	&$0.0^{+0.12}_{h}$	&0.77
&$2.57^{+1.25}_{-0.60}$	&$1.96^{h}_{-1.16}$	&$9.68^{+1.29}_{-9.68}$	&0.50\\
	      &$2.02 \pm {0.16}$	&$0.26^{+0.02}_{-0.43}$ &fixed 
			&--   &0.48 & & & & & & & &\\

1415.2$+$4403 &$2.97^{+0.50}_{-0.43}$	&$0.146^{+0.03}_{-0.04}$	&$1.72^{+0.96}_{-0.72}$	&1.17 &1.13
&$123^{+10}_{-12}$	&$0.98^{+0.79}_{-0.17}$	&$(1.05^{+3594}_{h})\times 10^{-4}$	&1.57
&$1.04^{+0.24}_{-0.14}$	&$1.91^{h}_{-0.37}$	&$16.5^{+2.7}_{-16.5}$	&1.31\\
	      &$2.70^{+0.22}_{-0.20}$	&$1.52^{+0.25}_{-0.48}$ &fixed 	
		&--   &1.12 & & & & & & & &\\

1416.2$+$1136 &$2.98^{+1.11}_{-0.69}$	&$0.14^{+0.06}_{-0.03}$	&$2.61^{+3.11}_{-1.78}$	&1.81 &0.90	
&$145 \pm 14$ &$0.79^{+0.23}_{-0.14}$	&$0.0^{+0.64}_{h}$	&0.95	
&$1.23^{+0.33}_{-0.52}$	&$1.94^{h}_{-0.96}$	&$15.4^{+2.83}_{-15.4}$ &0.89\\
&$2.68^{+0.27}_{-0.24}$	&$1.54^{+0.31}_{-0.54}$ &fixed 	
		&--   &0.86 & & & & & & & &\\

1420.0$+$0625 &$2.21^{+0.38}_{-0.34}$	&$0.52^{+0.06}_{-0.08}$  &$3.84^{+2.23}_{-1.58}$	
&2.18 &0.82
&$259^{+33}_{-27}$	&$1.42 \pm 0.15$ &$0.0^{+0.4}_{h}$	 &0.88
&$2.92^{+5.09}_{-1.69}$ &$1.03^{h}_{h}$ &$7.31^{+6.17}_{-7.31}$ &0.80\\
&$1.79^{+0.17}_{-0.19}$	&$4.66^{+0.46}_{-0.51}$ &fixed 	
	&--   &0.82 & & & & & & & &\\

1535.0$+$2336 &$0.70^{+0.72}_{-1.05}$	&$0.14^{+0.04}_{-0.04}$	&$0.16^{+2.37}_{-0.16}$	&4.26 &1.67
&$1150^{h}_{h}$	&$3.16^{h}_{h}$	&$0.0^{+20.58}_{h}$	&1.74
&$64^{h}_{-61.22}$	&$0.00^{h}_{h}$	&$1.16^{+4.33}_{-1.16}$	&1.76\\
	      &$0.47^{+1.72}_{-1.68}$	&$1.40 \pm 0.66$ &fixed 	
		&--   &1.66 & & & & & & & &\\

\hline
absorbed sample	&$2.03^{+1.21}_{-1.02}$	
&$0.90^{+0.76}_{-0.90}$ &$23.31^{+20.58}_{-15.33}$ &1.49	&0.46	
&$372^{+86}_{-75}$ &$1.99^{+0.35}_{-0.22}$ &$6.53^{+7.73}_{-5.07}$	&0.49
&$64.0^{h}_{-34.63}$	&$0.00^{+1.09}_{h}$	&$2.28^{+3.35}_{-2.28}$	&0.96\\
&$0.48^{+0.21}_{-0.24}$	&$0.46^{+0.30}_{-0.38}$	 &(1.49)fixed &-- &0.72	& & & & & & & &\\
unabsorbed sample	&$1.85 \pm 0.1$ &--$^i$ &$1.43^{+0.28}_{-0.15}$ &1.45 &1.27
&208  &--$^i$ &0. &2.54
&$3.49^{+0.63}_{-0.53}$ &$1.79^{h}_{-0.44}$  &--$^i$  &1.41\\
	&$1.86 \pm 0.04$ &--$^i$ &1.45(fixed) &-- &1.26  & & & &  & & & &
\enddata
\label{blanks_tab_fits}
\tablenotetext{a}{Photon index}
\tablenotetext{b}{normalizations at 1~keV in $10^{-4}$  for the power law and the Raymond-Smith models and in $10^{-6}$ ph cm$^{-2}$ s$^{-1}$ keV$^{-1}$ for the blackbody model}
\tablenotetext{c}{Hydrogen  column density for an absorbed power law in $10^{20}$ atoms cm$^{-2}$}
\tablenotetext{d}{Galactic hydrogen column density (Stark et al. 1992)}
\tablenotetext{e}{Reduced $\chi ^2$}
\tablenotetext{f}{Temperature in eV}
\tablenotetext{g}{Temperature in KeV}
\tablenotetext{h}{value pegged to the hard limit}
\tablenotetext{i}{We are simultaneously fitting two datasets
(i.e. the sources observed before and after the change in PSPC gain) and we have two different normalizations}
\end{deluxetable}
\end{landscape}

\newpage

\begin{landscape}
\begin{deluxetable}{c c c c c c c c }
\tablewidth{8.4in}
\footnotesize
\tablecaption{X-ray (absorbed) fluxes for the 16 \bfs .}
\tablehead{
\colhead{Source name} 
& \colhead{Class.}
& \colhead{$F_{X}$\tablenotemark{a}} 
& \colhead{$F_{X}$\tablenotemark{a}} 
& \colhead{$F_{X}$\tablenotemark{a}} 
& \colhead{$F_{{\rm WGA}}$\tablenotemark{b}} 
& \colhead{$F_{X}$/$F_{{\rm WGA}}$}
& \colhead{$F_{X}$/$F_{V}$\tablenotemark{c}}\\
\colhead{(1WGA~J)}
&
& \colhead {(0.3-3.5 {\rm keV})}
& \colhead {(0.5-2.0 {\rm keV})}
& \colhead {(0.05-2.5 {\rm keV})}
& \colhead {(0.05-2.5 {\rm keV})}
& \colhead {(0.05-2.5 {\rm keV})}
&
}
\startdata
0221.1$+$1958	&High $N_{H{\rm Gal.}}$	&2.28	&1.47	&1.91	&1.98	&0.96	&$>$21.3\\
0432.4$+$1723	&High $N_{H{\rm Gal.}}$	&4.12	&1.80	&2.63	&1.69	&1.56	&$>$38.5\\
0951.4$+$3916	&Absorbed	&1.39	&0.99 	&1.19	&2.59	&0.46	&$>$13.0\\
1103.5$+$2459\tablenotemark{d}	&Unabsorbed	&0.64	&0.37	&0.64	&1.01	&0.63	&$>$3.4\\
1216.9$+$3743	&Absorbed	&1.45	&1.22 	&1.40	&2.80	&0.50	&$>$13.5\\
1220.3$+$0641	&Unabsorbed	&8.22	&4.74	&7.90	&14.39	&0.53	&102.6\\
1220.6$+$3347	&Unabsorbed	&1.30	&0.79 	&1.40	&2.00	&0.70	&14.6\\
1226.9$+$3332	&Unabsorbed	&4.44	&2.43 	&3.61	&3.61	&1.00	&$>$41.4\\
1233.3$+$6910	&Unabsorbed	&4.11	&2.41	&4.15	&2.99	&1.39	&$>$38.4\\
1243.6$+$3204	&Absorbed	&1.04	&0.83 	&0.96	&1.28	&0.75	&$>$9.7\\
1340.1$+$2743	&Unabsorbed	&8.94	&5.60	&8.70	&10.14	&0.86	&$>$83.4\\
1412.3$+$4355\tablenotemark{d}	&Unabsorbed	&0.97	&0.56	&0.97	&1.63	&0.60	&$>$9.0\\
1415.2$+$4403	&Unabsorbed	&0.64	&0.32	&0.81	&1.21	&0.67	&$>$6.0\\
1416.2$+$1136	&Unabsorbed	&0.59	&0.30	&0.69	&1.11	&0.62	&10.7\\
1420.0$+$0625	&Unabsorbed	&1.68	&1.05	&1.53	&2.80	&0.53	&$>$15.7\\
1535.0$+$2336	&High $N_{H{\rm Gal.}}$	&0.83	&0.36	&0.55	&1.04	&0.53	&10.2\\
\enddata
\label{blanks_fx}
\tablenotetext{a}{in units of $10^{-13}$ \cgs \/ and computed 
 assuming the best fit absorbed power law model 
with free absorption.}
\tablenotetext{b}{in units of $10^{-13}$ \cgs \/ and computed using 
a constant correction factor of $1.5 \times 10^{-11}$ \cgs .}
\tablenotetext{c}{$F_{X}$ in the 0.3-3.5 {\rm keV} band divided by  $F_{V}$}
\tablenotetext{d}{Since the fit with an absorbed power law model with free 
absorption gives a $\sim 4$ times lower than Galactic absorbing column, we 
froze the absorption to the Galactic value.}
\end{deluxetable}				   
\end{landscape}

\newpage 
\begin{deluxetable}{l c c c c c}
\footnotesize
\tablewidth{7.2in} 
\tablecaption{Sources detected in the RASS}
\tablehead{
\colhead{Source name}
& \colhead{Offset\tablenotemark{a}}
& \colhead{RASS}
& \colhead{Pointed Obs.}
& \colhead{RASS}
& \colhead{Pointed Obs.}\\
\colhead{(1WGA~J)} 
& \colhead{($^{\prime \prime}$)}
& \colhead{count rate\tablenotemark{b}}
& \colhead{count rate\tablenotemark{c}}
& \colhead{Obs. Date}
& \colhead{Obs. Date}
}
\startdata
1220.3$+$0641	&10.5	&$8 \pm 2$		&$9.88 \pm 0.64$	&not available	&12-1991\\
1220.6$+$3347	&13.9	&$3.12 \pm 1.06$	&$1.92 \pm 0.14$	&12-1990	&06-1992 and 5-1993\\
1233.3$+$6910	&15.6	&$2.77 \pm 0.83$	&$4.48 \pm 0.24$	&11-1990	&04-1991\\
\enddata
\label{tab_rass}
\tablenotetext{a}{Offset calculated from the positions in Table~\ref{blanks_tab_16src}}
\tablenotetext{b}{In units of $10^{-2}$ counts s$^{-1}$.}
\tablenotetext{c}{Full band PSPC count rate in units of $10^{-2}$ counts s$^{-1}$.}
\end{deluxetable}


\begin{deluxetable}{c c c c c }
\footnotesize
\tablewidth{4.8in}
\tablecaption{{\em ASCA} serendipitous observations of {\em ROSAT} \bfs .}
\tablehead{
\colhead{Source name}
& \colhead{ROR}
& \colhead{Date}
& \colhead{Exposure}
& \colhead{Count Rate\tablenotemark{a}}\\
\colhead{(1WGA~J)}
& 
&
&\colhead{(ks)\tablenotemark{b}}
&
}
\startdata
1220.3$+$0641	&74074000	&12-25-1995	&46	&$21.4$\tablenotemark{c}\\
\hline
1226.9$+$3332	&76006000	&05-24-1998	&44	&$3.83 \pm 0.48$\tablenotemark{d}\\
		&78009000  	&05-25-2000 	&77	&\tablenotemark{e}\\
		&78009001  	&05-26-2000	&158 	&$4.43 \pm 0.30$\tablenotemark{d}\\
		&78009002  	&05-28-2000 	&133	&\tablenotemark{e}\\
		&78009003  	&05-30-2000	&188	&$3.80 \pm 0.32$\tablenotemark{d}\\
\hline
1415.2$+$4403	&74075000	&12-08-1996	&76	&1.55\tablenotemark{f}\\
\hline
1535.0$+$2336	&60035000	&07-26-1993	&68	&$0.80 \pm 0.26$\\
\enddata
\label{blanks_tab_asca}
\tablenotetext{a}{GIS2$+$GIS3 count rate in units of $10^{-3}$ counts s$^{-1}$ in the full {\em ASCA} band}
\tablenotetext{b}{GIS2$+$GIS3 archival net exposure time}
\tablenotetext{c}{AMSS (Ueda et al. 2002) count rate in the 0.7-7.0 keV band}
\tablenotetext{d}{The source falls close to the border of the useful area of the GIS detectors; the
value is a lower limit of the source count rate}
\tablenotetext{e}{The source falls on the border of the useful area of the GIS detectors}
\tablenotetext{f}{3$\sigma$ upper limit in a 24 pixel ($=6^{\prime}$) radius circle}
\end{deluxetable}

\newpage
\begin{deluxetable}{c c c c c c}
\tablewidth{7.0in}
\tablecaption{The radio counterparts to the \bfs .}
\tablehead{
\colhead{Source}
&\colhead{NVSS coord.}
&\colhead{FIRST coord.}
&\colhead{NVSS flux\tablenotemark{a}}
&\colhead{FIRST flux\tablenotemark{a}}\\
\colhead{(1WGA~J)}
&\colhead{Ra, Dec (J2000)}
&\colhead{Ra, Dec (J2000)}
&\colhead{(mJy)}
&\colhead{(mJy)}
}
\startdata
0221.1$+$1958	&02 21 09.25, 19 58 07.2 	& not covered yet		&$7.9 \pm 0.5$	& --\\
0432.4$+$1723	&04 32 30.53, 17 23 41.0 	& not covered yet		&$3.2 \pm 0.6$	& --\\
1226.9$+$3332	&12 26 58.32, 33 32 44.3 	&12 26 58.186,  33 32 48.63	&$4.9 \pm 0.6$	&3.61\\
1340.1$+$2743	&13 40 10.93, 27 43 45.7 	&13 40 10.853,  27 43 46.96 	&$4.6 \pm 0.5$	&3.75\\
\enddata
\label{blanks_tab_radio}
\tablenotetext{a}{@ 1.4 GHz}
\end{deluxetable}

\newpage

\begin{deluxetable}{c c c c c c }
\footnotesize
\tablewidth{4.8in}
\tablecaption{Classification of the blanks.}
\tablehead{
\colhead{Source name}
& \colhead{Identification}
& \colhead{z range}
& \colhead{E\tablenotemark{a}}
& \colhead{O\tablenotemark{a}} 
& \colhead{O-E\tablenotemark{b}}\\
\colhead{(1WGA~J)}
& &&&&
}
\startdata
1340.1$+$2743	 &BL Lac	&	 	&- 	&- 		&-\\
\hline
0221.1$+$1958	 &Cluster	&0.45	 	&- 	&- 	 	&-\\
0432.4$+$1723	 &Cluster?	&$\sim 0.5-1.0$	&20.0	&- 	 	&$>1.5$\\
1103.5$+$2459	 &Cluster?	&$\sim 1$ 	&18.2 &-  		&$>3.3$\\
1226.9$+$3332	 &Cluster	&0.89	 	&- 	&- 		&-\\
\hline
0951.4$+$3916	 &AGN?		&	 	&19.1 	&- 		&$>2.4$	\\	
1220.3$+$0641	 &AGN?		&	 	&18.58 	&21.81 		&3.23\\
1233.3$+$6910	 &AGN?		&	 	&- 	&- 		&-\\
1416.2$+$1136	 &AGN?		&	 	&19.5    &22.22 	 &2.72\\
1412.3$+$4355	 &AGN		&0.59	 	&18.73 	&- 		&$>2.77$\\
1415.2$+$4403	 &AGN		&0.56	 	&- 	&- 		&-	\\
1535.0$+$2336	 &AGN?		&	 	&19.6 	&21.8 	 	&2.2\\
\hline
1216.9$+$3743	 &X-ray binary? &	 	&-  	&- 	 	&-\\	
1243.6$+$3204	 &X-ray binary?	&	 	&-  	&- 		&-\\	
\hline	
1220.6$+$3347	 &Unknown	&       	&20.9 	&21.7 		&0.8\\	
1420.0$+$0625	 &Unknown	&	 	&- 	&- 		&-\\
\enddata
\label{blanks_tab_ids}

\tablenotetext{a}{Palomar data}
\tablenotetext{b}{when only the O or E magnitude is available we computed lower/upper 
limits on O-E
assuming O=21.5 and E=19.5 as Palomar limits}
\end{deluxetable}

\newpage

\newpage

\end{document}